\documentclass[11pt,a4paper]{article}
 \pdfoutput=1

\usepackage{amsmath,amssymb}
\usepackage{graphicx}
\usepackage{dcolumn}
\usepackage{bm,color}
\usepackage{hyperref}
\usepackage{accents}
\usepackage{amssymb,float}
\usepackage{amsmath}
\usepackage{multirow}
\usepackage{siunitx}
\usepackage{tabularx}
\usepackage{booktabs}
\usepackage{authblk}
\usepackage{url}
\usepackage{geometry}
\usepackage{caption}
\DeclareSIUnit\parsec{pc} 
\hypersetup{
    colorlinks=true,
    linkcolor=red,
    filecolor=magenta,      
    citecolor=blue
}

\newcommand{\udt}[3]{#1^{#2}_{\phantom{#2}#3}}

\newcommand{\dut}[3]{#1_{#2}^{\phantom{#2}#3}}

\newcommand{\lc}[1]{\accentset{\circ}{#1}}

\title{Constraints on \texorpdfstring{$f(T)$}{f(T)} Cosmology with Pantheon+}

\author[1,2]{Rebecca Briffa, \thanks{\href{rebecca.briffa.16@um.edu.mt}{rebecca.briffa.16@um.edu.mt}}}
\author[3]{Celia Escamilla-Rivera,\thanks{\href{celia.escamilla@nucleares.unam.mx}{celia.escamilla@nucleares.unam.mx}}}
\author[1,2]{Jackson Levi Said, \thanks{\href{jackson.said@um.edu.mt}{jackson.said@um.edu.mt}}}
\author[1,2]{and Jurgen Mifsud \thanks{\href{jurgen.mifsud@um.edu.mt}{jurgen.mifsud@um.edu.mt}}}

\affil[1]{Institute of Space Sciences and Astronomy, University of Malta, Malta, MSD 2080}
\affil[2]{Department of Physics, University of Malta, Malta}
\affil[3]{Instituto de Ciencias Nucleares, Universidad Nacional Aut\'{o}noma de M\'{e}xico, Circuito Exterior C.U., A.P. 70-543, M\'exico D.F. 04510, M\'{e}xico}

\date{\today}

\begin{document}

\maketitle
\abstract{
$f(T)$ cosmology has shown promise in explaining aspects of cosmic evolution. In this work, we analyze constraints on leading models of $f(T)$ gravity in the context of the recently released Pantheon+ data set, together with comparisons with previous releases. We also consider other late time data sets including cosmic chronometers and baryonic acoustic oscillation data. Our main result is that we find that the different $f(T)$ models under investigation connect to a variety of Hubble constant, which may help alleviate the cosmic tension on this parameter.
}


\section{\label{sec:intro}Introduction}

$\Lambda$CDM model has been supported by unprecedented observational evidence at all cosmic scales for several decades as the standard model of cosmology \cite{Misner:1974qy,Clifton:2011jh} with cold dark matter (CDM) acting as a stabilizing agent in galaxies \cite{Baudis:2016qwx,Bertone:2004pz}, and dark energy realized through the cosmological constant \cite{Peebles:2002gy,Copeland:2006wr}. However, despite great efforts, internal consistency issues persist in the cosmological constant description of cosmology \cite{Weinberg:1988cp}, while direct measurements of any dark matter particles remains elusive \cite{Gaitskell:2004gd}. More recently, the effectiveness of the $\Lambda$CDM model has come into question with the appearance of statistical tensions between some cosmic surveys which has taken the form of the so-called $H_0$ tension \cite{DiValentino:2020zio}. One perspective of the discrepancy is between model-independent measurements of the Hubble parameter at late times \cite{Riess:2019cxk,Wong:2019kwg} and the predictive power of the $\Lambda$CDM model using early time measurements \cite{Aghanim:2018eyx,Ade:2015xua}, or it may be an artifact of some types of measurements \cite{Riess:2020sih,Pesce:2020xfe,deJaeger:2020zpb}. Ultimately, the issue may even take new types of measurements to fully resolve the possible extent of the tension such as through gravitational wave standard sirens \cite{Baker:2019nia,2017arXiv170200786A,Barack:2018yly}.

The growing pressure on the $\Lambda$CDM model \cite{Bernal:2016gxb,DiValentino:2020zio,DiValentino:2021izs} has prompted a re-exploration of possible alternatives to its fundamental formulation \cite{Sotiriou:2008rp,Clifton:2011jh,CANTATA:2021ktz}. These alternatives are largely built on correction terms to the Einstein-Hilbert action where the gravitational field continues to be communicated by the curvature associated with the Levi-Civita connection \cite{misner1973gravitation,nakahara2003geometry}. On the other hand, there is a growing body of work that considers torsion rather than curvature as the mode by which gravity is exhibited on manifolds \cite{Bahamonde:2021gfp,Aldrovandi:2013wha,Cai:2015emx,Krssak:2018ywd}. Teleparallel gravity (TG) embodies the breadth of theories in which gravity is based on the torsion associated with the teleparallel connection. The teleparallel connection is curvature-less and satisfies metricity, and so all measures of curvature identically vanish irrespective of the components of the metric. One consequence of this exchange of connections is that the Ricci scalar, as calculated using the curvature-less teleparallel connection, will vanish, i.e. $R=0$, while its regular form $\lc{R}$ (over-circles represent quantities calculated with the Levi-Civita connection) will naturally remain arbitrary in value. Analogous to the Ricci scalar, TG produces a torsion scalar $T$ which is equal to the regular Ricci scalar up to a total divergence term $B$, making the action based on the linear form of the torsion scalar dynamically equivalent to general relativity (GR), also called the teleparallel equivalent of general relativity (TEGR).

As in curvature-based gravity models, TEGR can be modified to form different extensions to standard gravity. In fact, TEGR can be directly generalized to form $f(T)$ gravity \cite{Ferraro:2006jd,Ferraro:2008ey,Bengochea:2008gz,Linder:2010py,Chen:2010va,Bahamonde:2019zea,Paliathanasis:2017htk,Farrugia:2020fcu,Bahamonde:2021srr,Bahamonde:2020bbc}, which is a second order gravitational theory that has shown promise in meeting some observational challenges in both the cosmological and astrophysical sectors \cite{Cai:2015emx,Farrugia:2016qqe,Finch:2018gkh,Farrugia:2016xcw,Iorio:2012cm,Deng:2018ncg}. For instance, in Refs.~\cite{Nesseris:2013jea,Anagnostopoulos:2019miu} both expansion and growth data sets are used to constrain prominent models within $f(T)$ gravity. $f(T)$ gravity has also been explored using the CMB power spectrum in Ref.~\cite{Nunes:2018evm} for a power-law model. While in Ref.~\cite{Benetti:2020hxp} big bang nucleosynthesis data was used to constrain other models.

In addition to the public data sets, survey results can also be used in conjunction as priors to further analyze their consistency with said data sets. For instance, in Ref.~\cite{Riess:2019cxk} the SH0ES Team estimates the Hubble constant to be $73.30 \pm 1.04 \,{\rm km\, s}^{-1} {\rm Mpc}^{-1}$ which was reported using Supernova Type Ia events (SNIa), while the H0LiCOW Collaboration's \cite{Wong:2019kwg} measurement of $73.3^{+1.7}_{-1.8} \,{\rm km\, s}^{-1} {\rm Mpc}^{-1}$ relies on strong lensing from quasars. One of the lowest reported local values of the Hubble constant comes from measurements based on using the tip of the red giant branch (TRGB) as a standard candle with $H_0=69.8 \pm 1.9 \,{\rm km\, s}^{-1} {\rm Mpc}^{-1}$ as reported in Ref.~\cite{Freedman:2019jwv}. Together with cosmic chronometer, SNIa and baryonic acoustic oscialltions, the impact of these priors on the most studied $f(T)$ gravity models was recently studied in Ref.~\cite{Briffa:2021nxg}. The SNIa data set used in this study relied on the Pantheon release (PN) which is a compilation of 1048 SNIa relative luminosity distance measurements spanning the redshift range of $0.01<z<2.3$ \cite{Scolnic:2017caz}. More recently the Pantheon+ data ($\mathrm{PN}^+\,\&\,$SH0ES) set has been released which builds on the Pantheon data set and features 1701 events with a much higher concentration of data points at lower redshift bins \cite{Brout:2021mpj,Riess:2021jrx,Scolnic:2021amr}. This drastic increase in data points may yield much stronger constraints on cosmological models beyond $\Lambda$CDM such as $f(T)$ gravity models.

In the present work, we perform constraint analyses using $\mathrm{PN}^+\,\&\,$SH0ES for the most promising $f(T)$ gravity models which we then compare with previous studies using other data sets. This lets us compare the impact of $\mathrm{PN}^+\,\&\,$SH0ES with the PN data set. We start by first reviewing some technical details of TG in Sec.~\ref{sec:TG}, which is then followed by a description of the data sets being used in Sec.~\ref{Sec3:methodology}. Our main results can be found in Sec.~\ref{sec:results} where we constrain our $f(T)$ gravity models using these data sets. We also present a comparison of our analyses with the standard model of cosmology in Sec.~\ref{sec:model_comparison}. Finally, we summarize our main results and discuss possible future work in Sec.~\ref{sec:conc}.

\section{Teleparallel Cosmology} \label{sec:TG}

TG is sourced by the exchange of the curvature-based Levi-Civita connection $\udt{\lc{\Gamma}}{\sigma}{\mu\nu}$ (over-circles are used throughout to denote quantities determined using the Levi-Civita connection) with the teleparallel connection $\udt{\Gamma}{\sigma}{\mu\nu}$ \cite{Hayashi:1979qx,Aldrovandi:2013wha,Bahamonde:2021gfp}. The curvature-less nature of the teleparallel connection means that all curvature-based geometric bodies will vanish identically (the regular curvature-based quantities remain arbitrary when calculated using the Levi-Civita connection) when calculated using this connection, and so new quantities are needed to build gravitational theories \cite{Krssak:2018ywd,Cai:2015emx,Aldrovandi:2013wha}.

Curvature-based gravitational models are largely built on the metric tensor, while TG is most directly expressed through the tetrad $\udt{e}{A}{\mu}$ (and its inverses $\dut{E}{A}{\mu}$) and spin connection $\udt{\omega}{A}{B\mu}$. The tetrad $\udt{e}{A}{\mu}$ builds up to the metric through
\begin{align}\label{metric_tetrad_rel}
    g_{\mu\nu} = \udt{e}{A}{\mu}\udt{e}{B}{\nu}\eta_{AB}\,,& &\eta_{AB} = \dut{E}{A}{\mu}\dut{E}{B}{\nu}g_{\mu\nu}\,,
\end{align}
where Latin indices represent coordinates on the tangent space while Greek indices represent coordinates on the general manifold \cite{Cai:2015emx}. In GR, the appearance of tetrads is largely suppressed since the tetrad is not the only non-inertial variable in that description of gravity. As with the metric, the tetrad observes orthogonality conditions, namely
\begin{align}
    \udt{e}{A}{\mu}\dut{E}{B}{\mu}=\delta^A_B\,,&  &\udt{e}{A}{\mu}\dut{E}{A}{\nu}=\delta^{\nu}_{\mu}\,,
\end{align}
for internal consistency. The spin connection $\udt{\omega}{A}{B\mu}$ is a flat spin connection and is responsible for incorporating the local Lorentz transformation invariance into the equations of motion, which arises due to the appearance of the tangent space indices.

The tetrad and spin connection define the teleparallel connection through \cite{Weitzenbock1923,Krssak:2018ywd}
\begin{equation}
    \udt{\Gamma}{\sigma}{\nu\mu} := \dut{E}{A}{\sigma}\left(\partial_{\mu}\udt{e}{A}{\nu} + \udt{\omega}{A}{B\mu}\udt{e}{B}{\nu}\right)\,.
\end{equation}
Together, the tetrad and spin connection represent the gravitational and local degrees of freedom of the system, and retain the diffeomorphism and local Lorentz invariance of the equations of motion. Analogous to the way in which the Levi-Civita connection builds up to the Riemann tensor, the torsion tensor can be constructed from the teleparallel connection as \cite{Hayashi:1979qx}
\begin{equation}\label{eq:tor_scal_def}
    \udt{T}{\sigma}{\mu\nu} := 2\udt{\Gamma}{\sigma}{[\nu\mu]}\,,
\end{equation}
where square brackets denote an antisymmetric operator. Considering a particular contraction of the torsion tensor, a torsion scalar can be put together \cite{Krssak:2018ywd,Cai:2015emx,Aldrovandi:2013wha,Bahamonde:2021gfp}
\begin{equation}
    T:=\frac{1}{4}\udt{T}{\alpha}{\mu\nu}\dut{T}{\alpha}{\mu\nu} + \frac{1}{2}\udt{T}{\alpha}{\mu\nu}\udt{T}{\nu\mu}{\alpha} - \udt{T}{\alpha}{\mu\alpha}\udt{T}{\beta\mu}{\beta}\,,
\end{equation}
which is equal to the curvature-based Ricci scalar up to a total divergence term. Thus, the TEGR action is represented by a linear Lagrangian form of the torsion scalar since \cite{Bahamonde:2015zma,Farrugia:2016qqe}
\begin{equation}\label{LC_TG_conn}
    R=\lc{R} + T - B = 0\,,
\end{equation}
where $R\equiv0$ since the teleparallel connection is curvature-less, while $\lc{R} \neq 0$ since this is determined using the Levi-Civita connection, while the boundary term $B$ is a total divergence term. Thus, the Einstein-Hilbert action is dynamically equivalent to the representation of a linear torsion scalar which guarantees identical equations of motion for the two actions.

As curvature-based gravity, modification of TEGR can be designed and explored, with the most direct being the arbitrary generalization of the TEGR Lagrangian to $f(T)$ gravity, which we parameterize as $f(T) = -T + \mathcal{F}(T)$ gravity by raising the TEGR action \cite{Ferraro:2006jd,Ferraro:2008ey,Bengochea:2008gz,Linder:2010py,Chen:2010va,RezaeiAkbarieh:2018ijw} through the action
\begin{equation}\label{f_T_ext_Lagran}
    \mathcal{S}_{\mathcal{F}(T)}^{} =  \frac{1}{2\kappa^2}\int \mathrm{d}^4 x\; e\left(-T + \mathcal{F}(T)\right) + \int \mathrm{d}^4 x\; e\mathcal{L}_{\text{m}}\,,
\end{equation}
where $\kappa^2=8\pi G$, $\mathcal{L}_{\text{m}}$ is the matter Lagrangian, and $e=\det\left(\udt{e}{a}{\mu}\right)=\sqrt{-g}$ is the tetrad determinant. A healthy TEGR exists for the case when $\mathcal{F}(T) \rightarrow 0$ and the $\Lambda$CDM model is obtained when this functional tends to a constant $\Lambda$ value. The $\mathcal{F}(T)$ equations of motion are particular in that they are generaically second order in nature and so do not exhibit any Gauss-Ostrogadsky ghosts \cite{Aldrovandi:2013wha}. Indeed, the field equations can be written through
\begin{align}\label{ft_FEs}
    \dut{W}{a}{\mu} := e^{-1} &\partial_{\nu}\left(e\dut{E}{a}{\rho}\dut{S}{\rho}{\mu\nu}\right)\left(-1 + \mathcal{F}_T\right) - \dut{E}{a}{\lambda} \udt{T}{\rho}{\nu\lambda}\dut{S}{\rho}{\nu\mu} \left(-1 + \mathcal{F}_T\right) + \frac{1}{4}\dut{E}{a}{\mu}\left(-T + \mathcal{F}(T)\right) \nonumber\\
    & + \dut{E}{a}{\rho}\dut{S}{\rho}{\mu\nu}\partial_{\nu}\left(T\right)\mathcal{F}_{TT}  + \dut{E}{b}{\lambda}\udt{\omega}{b}{a\nu}\dut{S}{\lambda}{\nu\mu}\left(-1 + \mathcal{F}_T\right) = \kappa^2 \dut{E}{a}{\rho} \dut{\Theta}{\rho}{\mu}\,,
\end{align}
where subscripts denote derivatives ($\mathcal{F}_T=\partial \mathcal{F}/\partial T$ and  $\mathcal{F}_{TT}=\partial^2 \mathcal{F}/\partial T^2$), and $\dut{\Theta}{\rho}{\nu}$ is the regular energy-momentum tensor. The individual tetrad and spin connection field equations are then represented by
\begin{equation}
    W_{(\mu\nu)} = \kappa^2 \Theta_{\mu\nu}\,, \quad \text{and} \quad W_{[\mu\nu]} = 0\,.
\end{equation}
For any metric, a unique tetrad-spin connection pairs exist that are compatible with a vanishing spin connection, called the Weitzenb\"{o}ck gauge \cite{Krssak:2018ywd,Bahamonde:2021gfp}. Here, $W_{[\mu\nu]}$ vanishes identically while continuing to satisfy the metric equations in Eq.~\eqref{metric_tetrad_rel}.

A flat homogeneous and isotropic cosmology is explored in this work through the tetrad \cite{Krssak:2015oua,Tamanini:2012hg}
\begin{equation}
    \udt{e}{A}{\mu} = \text{diag}\left(1,\,a(t),\,a(t),\,a(t)\right)\,,
\end{equation}
where $a(t)$ is the scale factor in cosmic time $t$, and which was shown to universally satisfy the Weitzenb\"{o}ck gauge conditions in Ref.~\cite{Hohmann:2019nat}. The regular flat Friedmann--Lema\^{i}tre--Robertson--Walker (FLRW) metric is reproduced using Eq.~\eqref{metric_tetrad_rel} so that the line element takes the regular form \cite{misner1973gravitation}
\begin{equation}\label{FLRW_metric}
     \mathrm{d}s^2 = \mathrm{d}t^2 - a^2(t) \left(\mathrm{d}x^2+\mathrm{d}y^2+\mathrm{d}z^2\right)\,,
\end{equation}
from which we can define the regular Hubble parameter as $H=\dot{a}/a$ where over-dots refer to derivatives with respect to cosmic time. Using Eqs.~(\ref{eq:tor_scal_def},\ref{LC_TG_conn}), it turns out that $T = -6 H^2$ and $B = -6\left(3H^2 + \dot{H}\right)$. Thus, the $f(T)$ gravity Friedmann equations can be written as \cite{Bahamonde:2021gfp}
\begin{align}
    H^2 + \frac{T}{3}\mathcal{F}_T - \frac{\mathcal{F}}{6} &= \frac{\kappa^2}{3}\rho\,,\label{eq:Friedmann_1}\\
    \dot{H}\left(1 - \mathcal{F}_T - 2T\mathcal{F}_{TT}\right) &= -\frac{\kappa^2}{2} \left(\rho + p \right)\label{eq:Friedmann_2}\,,
\end{align}
where we denote the energy density and pressure of the total matter sector by $\rho$ and $p$, respectively.


\section{Observational Data}
\label{Sec3:methodology}

In this study, we consider the most favourable $f(T)$ models and test them against different combinations of observational data sets. For each $f(T)$ model and data set combination, we perform an MCMC (Monte Carlo Markov Chain) analysis using the publicly available \textit{emcee} package available at Ref.~\cite{2013PASP..125..306F}. The MCMC sampler constrains the model and cosmological parameters by varying them in a range of conservative priors and exploring the posteriors of the parameter space. Therefore, for each parameter, we obtain its one- and two-dimensional distributions, where the one-dimensional distribution represents the parameters' posterior distribution whilst the two-dimensional one illustrates the covariance between two different parameters. These are complemented with their respective 1 and 2$\sigma$ confidence levels as shown in Sec.~\ref{sec:results}. In turn, this allows us to compare the different data sets and analyze the effects of $\mathrm{PN}^+\,\&\,$SH0ES with the PN data set.

We devote this section to present and describe the observational data which will be considered in the analyses below based on the MCMC analysis. Our baseline dataset consists of Hubble expansion data along with a SNIa.

\textbf{Cosmic Chronometers (CC)} - With regards to Hubble parameter data, we adopt thirty-one cosmic chronometer data points \cite{2014RAA....14.1221Z,Jimenez:2003iv,Moresco:2016mzx,Simon:2004tf,2012JCAP...08..006M,2010JCAP...02..008S,Moresco:2015cya}. This CC method involves spectroscopic dating techniques of passively-evolving galaxies, which enables us to directly obtain observational values of the Hubble functions at various redshifts up to, $z \lesssim2$. These measurements are independent of any cosmological model and the Cepheid distance scale, however, they are still associated with the modeling of the stellar ages, which is based on robust stellar population synthesis techniques. It involves the measurements of age difference between two passively-evolving galaxies at two redshifts. Therefore, $\Delta z/ \Delta t$ can be inferred from observations which in turn, makes it possible to compute $H(z) = -(1+z)^{-1} \Delta z/ \Delta t$. Thus, CCs were found to be more reliable than any other method that is based on the absolute age determination of galaxies \cite{Jimenez:2001gg}.

The corresponding $\chi^2_H$ estimator is given by 
\begin{equation}
    \chi^2_H = \sum^{31}_{i=1} \frac{\left(H(z_i,\Theta) - H_{\mathrm{obs}(z_i)}\right)^2}{\sigma^2_H(z_i)} \,,
\end{equation}
where $H(z_i, \Theta)$ are the theoretical Hubble parameter values at redshift $z_i$ with model parameters $\Theta$ whilst $H_{\mathrm{obs}}(z_i)$ are the corresponding Hubble data values at $z_i$ with observational error of $\sigma_H(z_i)$.

\textbf{Type Ia Supernovae Compilation} - The other baseline dataset used for our MCMC analyses includes information obtained from Type Ia supernovae. These supernovae occur in binary star systems and are valuable for cosmological analyses because of their uniform intrinsic brightness, which allows us to use them as standard candles to measure distances to distant galaxies. To be more specific, the difference between the observed apparent magnitude of an object, $m$, and its absolute magnitude, $M$ (which is a measure of its intrinsic brightness) is defined as the distance modulus. At redshift $z_i$, the distance modulus is given as
\begin{equation}
    \mu(z_i, \Theta) = m - M = 5 \log_{10}[D_L(z_i, \Theta)] + 25 \,,
\end{equation}
where $D_L(z_i, \Theta)$ is the luminosity distance defined as 
\begin{equation}
    D_L(z_i, \Theta) = c(1+z_i) \int_0^{z_i} \frac{dz'}{H(z', \Theta)} \,. 
\end{equation}
In addition, the apparent magnitude of each SNIa needs to be calibrated via an arbitrary fiducial absolute magnitude $M$ and thus, in the MCMC analyses, we can treat $M$ as a nuisance parameter by marginalizing over it. This is done by using theoretical models to predict the distance modulus for a given set of cosmological parameters and comparing these predictions to the observed values for the SNIa in the Pantheon catalog. The cosmological parameters are then constrained by minimizing a $\chi^2$ likelihood specified by \cite{SNLS:2011lii}, 
\begin{equation}
    \chi^2_{\mathrm{SN}} = (\Delta \mu(z_i), \Theta))^T C^{-1} (\Delta \mu(z_i), \Theta)) \,,
\end{equation}
where $(\Delta \mu(z_i), \Theta)) = (\mu(z_i), \Theta) - \mu(z_i)_{\mathrm{obs}}$ and $C$ is the corresponding covariance matrix which accounts for the statistical and systematic uncertainties. 

In this work, we use two SNIa data sets: the Pantheon (PN) \cite{Pan-STARRS1:2017jku} and Pantheon+ ($\mathrm{PN}^+\,\&\,$SH0ES) \cite{Scolnic:2021amr} compilations, which is a successor to the original Pantheon analysis. The main difference between the original Pantheon analysis and the Pantheon+ analysis in cosmology lies in the addition of new data sets to the latter. While the original Pantheon analysis used a compilation of 1048 supernovae type Ia (SNIa) samples to study the expansion history of the Universe, the Pantheon+ analysis includes an even larger number of 1701 SNIa samples. The term ``$\mathrm{PN}^+\,\&\,$SH0ES'' as referred to in the Pantheon+ analysis in Ref.~\cite{Brout:2022vxf}, incorporates the SH0ES Cepheid host distance anchors (R22 \cite{Riess:2021jrx}) in the likelihood which helps to break the degeneracy between the parameters $M$ and $H_0$ when analyzing SNIa alone. Additionally, the Pantheon+ analysis covers a wider redshift range of $0.01 < z < 2.5$, compared to the original Pantheon, which does not extend redshifts lower than $z < 0.01$. This expanded redshift range allows for an improved treatment of systematic uncertainties, resulting in better-constrained parameters as will be illustrated in Sec.~\ref{sec:results}

\textbf{Baryon Acoustic Oscillations} - We also consider a joint baryon acoustic oscillation (BAO) data set consisting of independent data points. This BAO data set includes measurements from the SDSS Main Galaxy Sample at $z_{\mathrm{eff}} = 0.15$ \cite{Ross:2014qpa}, the six-degree Field Galaxy Survey at $z_{\mathrm{eff}} = 0.106$ \cite{2011MNRAS.416.3017B}, and the BOSS DR11 quasar Lyman-alpha measurement at $z_{\mathrm{eff}} = 2.4$ \cite{Bourboux:2017cbm}. We also incorporate the angular diameter distances and $H(z)$ measurements of the SDSS-IV eBOSS DR14 quasar survey at $z_{\mathrm{eff}} = \{0.98, 1.23, 1.52, 1.94\}$ \cite{Zhao:2018gvb}, along with the SDSS-III BOSS DR12 consensus BAO measurements of the Hubble parameter and the corresponding comoving angular diameter distances at $z_{\mathrm{eff}} = \{0.38, 0.51, 0.61\}$ \cite{Alam:2016hwk}. For these two BAO data sets, we consider the full covariance matrix in our MCMC analyses. 

For the BAO datasets under consideration, we compute the Hubble distance $D_H(z)$, comoving angular diameter distance $D_M(z)$, and volume-average distance $D_V(z)$ using 
\begin{equation}
    D_H(z) = \frac{c}{H(z)}, D_M(z) = (1+z)D_A(z), D_V(z) = \left[(1+z)^2D_A(z)^2 \frac{c z}{H(z)}\right]^{1/3}
\end{equation}
respectively, where $D_A(z)=(1+z)^{-2}D_L(z)$ is the angular diameter distance. Using the reported BAO results, we calculate the corresponding combination of parameters $\mathcal{G}(z_i)=D_V(z_i)/r_s(z_d)$,
$r_s(z_d)/D_V(z_i),$
$D_H(z_i),$ 
$D_M(z_i)(r_{s,\mathrm{fid}}(z_d)/r_s(z_d)),$
$H(z_i)(r_s(z_d)/r_{s,\mathrm{fid}}(z_d)),$ \\ 
$D_A(z_i)(r_{s,\mathrm{fid}}(z_d)/r_s(z_d))$ 
for which the comoving sound horizon at the end of the baryon drag epoch at redshift $z_d\approx 1059.94$ \cite{Planck:2018vyg} is computed by
\begin{equation}
    r_s(z)=\int_z^\infty\frac{c_s(\tilde{z})}{H(\tilde{z})}\,\mathrm{d}z=\frac{1}{\sqrt{3}}\int_0^{1/(1+z)}\frac{\mathrm{d}a}{a^2H(a)\sqrt{1+\left[3\Omega_{b,0}/(4\Omega_{\gamma,0})\right]a}}\,,
\end{equation}
where we have adopted $\Omega_{b,0}=0.02242$ \cite{Planck:2018vyg}, $T_{0}=2.7255\,\mathrm{K}$ \cite{2009ApJ...707..916F}, and a fiducial value of $r_{s,\mathrm{fid}}(z_d)=147.78\,\mathrm{Mpc}$.

The corresponding $\chi^2$ for the BAO data is calculated using
\begin{equation}
\chi^2_{\text{BAO}}(\Theta) = \Delta G(z_i,\Theta)^T C_{\text{BAO}}^{-1}\Delta G(z_i,\Theta)
\end{equation}
where $\Delta G(z_i,\Theta) = G(z_i,\Theta)-G_{\text{obs}}(z_i)$and $C_{\text{BAO}}$ is the covariance matrix of all the considered BAO observations.


\section{Results} \label{sec:results}

In this section, we present and analyze the results following the methodology outlined in Sec.~\ref{Sec3:methodology} and using the observational data previously discussed. Each subsection focuses on the most promising models of $f(T)$, presenting contour plots of the constrained parameters with $1\sigma$ and $2\sigma$ uncertainties, along with corresponding tables with final results. These models have gained prominence in literature and are frequently studied due to their ability to mirror very well our cosmological history. In all tables and posterior plots, we include results of the Hubble constant $H_0$, the current matter density parameter $\Omega_{m,0}$ together with the model parameters. This will allow us to analyze how the different independent data sets and cosmological models impact the Hubble tension.  We also provide a brief discussion of the most noteworthy findings, highlighting the differences between the PN and $\mathrm{PN}^+\,\&\,$SH0ES.


\subsection{Power Law Model}

The power law model, henceforth referred to as $f_1$CDM, which was introduced by Bengochea and Ferraro in \cite{Bengochea:2008gz}, proposes an alternative explanation for the observed acceleration of the late-time Universe that does not involve dark energy. The model introduces a modification function $\mathcal{F}_1(T)$, which has a power law form with two constant parameters $\alpha_1$ and $p_1$ specified by
\begin{equation} \label{eq:PLM}
    \mathcal{F}_1(T) = \alpha_1(-T)^{p_1} \,,
\end{equation}
The constant $\alpha_1$ can be calculated using the Friedman equation Eq.~\eqref{eq:Friedmann_1} at current times 
\begin{equation}
    \alpha_1 = (6H_0^2)^{1-p_1} \frac{1- \Omega_{m,0} - \Omega_{r,0}}{1-2p_1} \,,
\end{equation}
where $\Omega_{m,0}$ and $\Omega_{r,0}$ are the density parameter for matter and radiation at current times, respectively. Thus, instead of introducing two new parameters as in the original equation, only one new model parameter, $p_1$, is required for the $f_1$CDM model, making it a more simpler and elegant model. The value of $p_1$ can be obtained by applying the MCMC analyses to observational data.

The Friedmann equation for the $f_1$CDM model can, therefore, be obtained by substituting the above equation in Eq.~\eqref{eq:Friedmann_1} such that
\begin{equation}\label{eq:PLM_FE}
    E^2(z) = \Omega_{m,0}(1+z)^3 + \Omega_{r,0}(1+z)^4 + (1-\Omega_{m,0} - \Omega_{r,0}) E^{2p_1}(z) \,.
\end{equation}
Here, the normalised Hubble parameter $E(z) = \frac{H(z)}{H_0}$ was applied. It is worth noting that for $p_1 = 0$, Eq.~\eqref{eq:PLM_FE} reduces to $\Lambda$CDM, whereas for $p_1 = 1$, the GR limit is recovered as the additional component in the Friedmann equation produces a rescaled gravitational constant term in the density parameters. The objective is to obtain values of $H_0$, $\Omega_{m,0}$, and $p_1$ that provide the best fit to the observational data using the MCMC analyses.

\begin{figure}
    \centering
    \includegraphics[width = \linewidth]{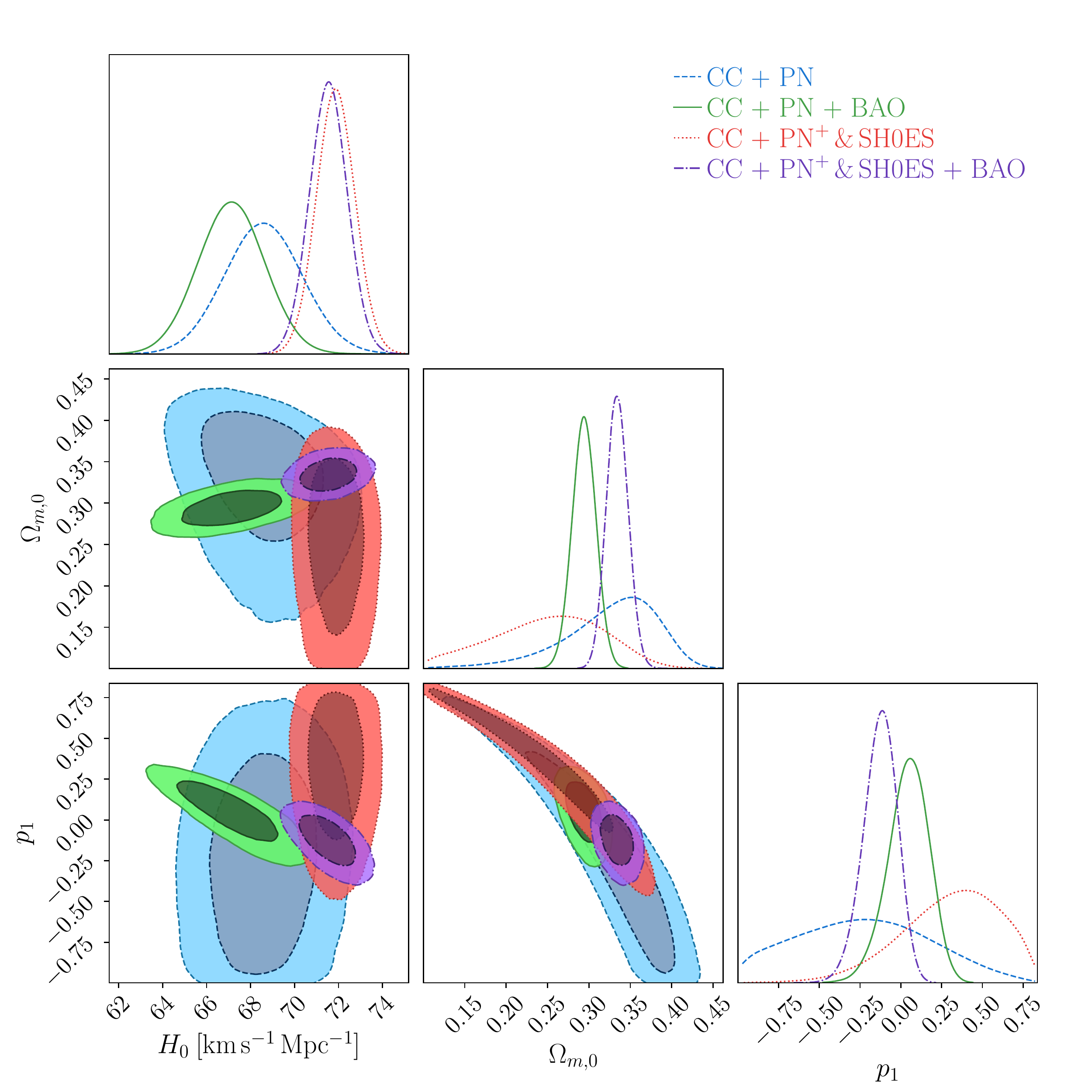}
    \caption{Confidence contours and posteriors for $f_1$CDM for the parameters $H_0$, $\Omega_{m,0}$ and $p_1$. The blue and green contours represent data set combinations that include PN data set, while the red and purple contours show combinations that also include the $\mathrm{PN}^+\,\&\,$SH0ES data sets.}
    \label{fig:PLM_CP+SB}
\end{figure}

The constraints on the specified parameters for $f_1$CDM model are shown in Fig.~\ref{fig:PLM_CP+SB}. The figure shows both the confidence regions and the posteriors for different combinations of observational data sets. Specifically, the figure shows the results for data sets that include either the PN catalog or the $\mathrm{PN}^+\,\&\,$SH0ES. Upon closer examination of the posteriors, it is evident that the parameters from the data set combinations that include $\mathrm{PN}^+\,\&\,$SH0ES exhibit tighter constraints, with the $H_0$ parameter showing notably improved precision. On the other hand, the contour plots for the CC+PN and CC+$\mathrm{PN}^+\,\&\,$SH0ES data set combinations display a degeneracy between the $H_0$ parameter and the $p_1$ parameter. However, once the BAO data set is included this degeneracy breaks and reveals an anti-correlation between the two parameters. It is noteworthy that the CC+$\mathrm{PN}^+\,\&\,$SH0ES data set combination shows a degeneracy between the $\Omega_{m,0}$ parameter and $H_0$, while for all data set combinations an anti-correlation is observed between the $p_1$ parameter and the $\Omega_{m,0}$ parameter. However, the strength of this anti-correlation is less pronounced for the data sets that include the BAO.

The precise values for the cosmological and model parameters, including the nuisance parameter $M$, for $f_1$CDM are shown in Table~\ref{tab:PLM}. It becomes clear that the values of $H_0$ for the data set combinations that include $\mathrm{PN}^+\,\&\,$SH0ES are relatively higher than their corresponding $H_0$ values. This finding is consistent with the high value of $H_0$ obtained by the SH0ES team (R22), which reports $H_0 = 73.30 \pm 1.04 \,{\rm km\, s}^{-1} {\rm Mpc}^{-1}$ \cite{Riess:2021jrx}. The results show that the highest values of $H_0$ are obtained for the CC+$\mathrm{PN}^+\,\&\,$SH0ES with a value of $H_0 = 71.88^{+ 0.87}_{- 0.89} \,{\rm km\, s}^{-1} {\rm Mpc}^{-1}$. Interestingly, in this scenario the $\Omega_{m,0}$ parameter reaches a minimum value, implying that most of the energy in the Universe appears as an effective dark energy, in line with the high value of $H_0$.

The inclusion of $\mathrm{PN}^+\,\&\,$SH0ES appears to better constrain the values of $p_1$, and this effect is even more pronounced with the addition of the BAO data. However, for the $\mathrm{PN}^+\,\&\,$SH0ES data set, the $p_1$ parameter is found to be within $1 \sigma$ of the corresponding $\Lambda$CDM value, whereas it moves to $2 \sigma$ for the $\mathrm{PN}^+\,\&\,$SH0ES combination.

The next section will provide a more detailed statistical analysis of these findings, including a comparison with the $\Lambda$CDM model.

\begin{table}
    \centering
    \small
    \caption{Results for the $f_1$CDM (Power law) model, where the first column lists the data sets used to constrain the parameters. The second to fourth columns display the constrained parameters, namely $H_0$, $\Omega_{m,0}$, and $p_1$, while the last column shows the nuisance parameter $M$.}
    \label{tab:PLM}
    \begin{tabular}{ccccc}
        \hline
		Data sets & $H_0\mathrm{\hspace{0.15cm}[km \hspace{0.1cm} s^{-1} \hspace{0.1cm}Mpc ^{-1}]}$ & $\Omega_{m,0}$ & $p_1$ & $M$ \\ 
		\hline
		CC + PN & $68.6^{+1.7}_{-1.8}$ & $0.352^{+0.042}_{-0.063}$ & $-0.22^{+0.41}_{-0.48}$ & $-19.390^{+0.052}_{-0.053}$ \\ 
		CC + PN + BAO & $67.1\pm 1.5$ & $0.294^{+0.015}_{-0.014}$ & $0.06^{+0.12}_{-0.13}$ & $-19.435\pm 0.044$ \\ 
		CC + $\mathrm{PN}^+\,\&\,$SH0ES & $71.88^{+0.87}_{-0.89}$ & $0.266^{+0.062}_{-0.076}$ & $0.40^{+0.28}_{-0.33}$ & $-19.295\pm 0.025$ \\ 
		CC + $\mathrm{PN}^+\,\&\,$SH0ES + BAO & $71.55^{+0.85}_{-0.86}$ & $0.334^{+0.014}_{-0.013}$ & $-0.113^{+0.098}_{-0.108}$ & $-19.309^{+0.024}_{-0.025}$ \\ 
		\hline
    \end{tabular}
\end{table}


\subsection{Linder Model}

The Linder model, henceforth referred to as $f_2$CDM, was specifically designed to account for the late-time acceleration of the Universe without the need for dark energy. This model incorporates a torsion scalar, $T$, and is described by the equation 
\begin{equation}
    \mathcal{F}_2 = \alpha_2 T_0 \left(1- \mathrm{Exp}\left[-p_2 \sqrt{T/T_0} \right] \right) \,,
\end{equation}
where $\alpha_2$ and $p_2$ are constants and $T_0$ represents the current value of the torsion scalar, that is $T|_{t=t_0}= -6H_0^2$. The constant $\alpha_2$ can be determined by evaluating the Friedmann equation at current times, which gives 
\begin{equation}
  \alpha_2 = \frac{1- \Omega_{m,0} - \Omega_{r,0}}{(1+b_2) e^{-b_2} - 1} \,.
\end{equation}
Therefore, the only new model parameter in the $f_2$CDM model is $p_2$. Using the above equations, the Friedmann equation for this model can be defined as
\begin{equation}
    E^2\left(z\right) = \Omega_{m_0} \left(1+z\right)^3 + \Omega_{r_0}\left(1+z\right)^4 + \frac{1 - \Omega_{m_0} - \Omega_{r_0}}{(p_2 + 1)e^{-b_2} - 1} \left[\left(1 + p_2 E(z)\right) \text{Exp}\left[-p_2 E(z)\right] - 1\right]\,. 
\end{equation}
This model can be reduced to $\Lambda$CDM when $p_2 \rightarrow \infty$. However, to ensure numerical stability, the analysis is performed for $1/b_2$, so that this limit becomes $1/b_2 \rightarrow 0^+$.

\begin{figure}
    \centering
    \includegraphics[width = \linewidth]{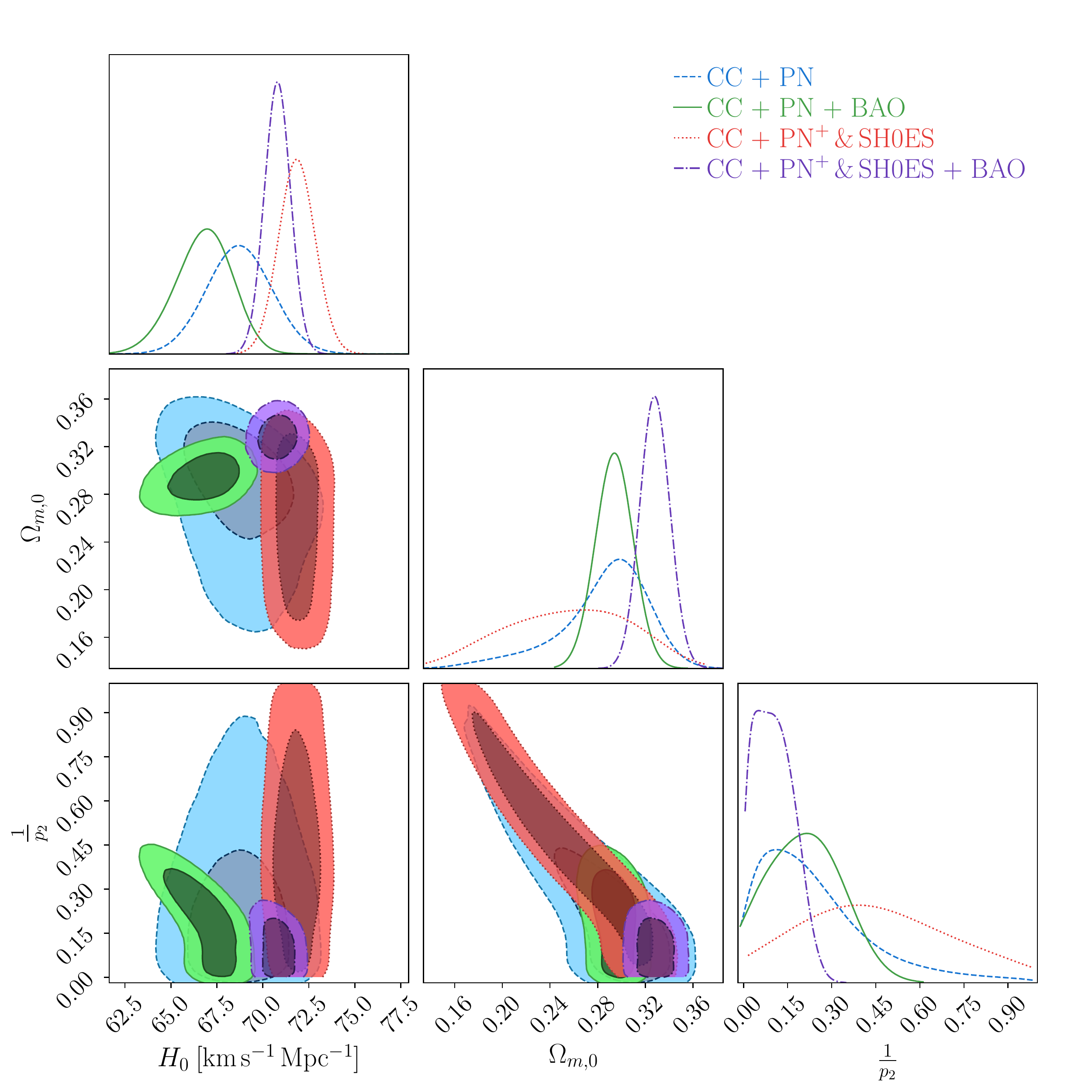}
    \caption{Confidence contours and posteriors for $f_2$CDM for the parameters $H_0$, $\Omega_{m,0}$ and $\frac{1}{p_2}$. The blue and green contours represent data set combinations that include PN data set, while the red and purple contours show combinations that also include the $\mathrm{PN}^+\,\&\,$SH0ES data sets.}
    \label{fig:LM_CP+SB}
\end{figure}

In Fig.~\ref{fig:LM_CP+SB}, the posterior and confidence levels of the constrained parameters for $f_2$CDM are displayed. The blue and green contours correspond to the combination of data sets that includes the PN sample, whereas the red and purple contours represent the combinations that consist of the $\mathrm{PN}^+\,\&\,$SH0ES samples. The $f_2$CDM model shows similar trends to the $f_1$CDM model, with tighter constraints for $\mathrm{PN}^+\,\&\,$SH0ES, particularly for the Hubble constant $H_0$, especially when the BAO data set is included. The CC+$\mathrm{PN}^+\,\&\,$SH0ES + BAO data set is the most constrained, indicating the highest precision. The anti-correlation between $\Omega_{m,0}$ and $\frac{1}{p_2}$ parameters remains evident in this model, particularly for data sets including the $\mathrm{PN}^+\,\&\,$SH0ES catalog.

Table~\ref{tab:LM} presents the exact numerical values of the parameters shown in Fig.~\ref{fig:LM_CP+SB}, including the nuisance parameter $M$. The results show that the estimated values of $H_0$ are comparable to those obtained in the $f_1$CDM model. However, as the $f_2$CDM model is specifically designed to predict an accelerating Universe in the late-time regime, the inferred values of the matter density parameter $\Omega_{m,0}$ are slightly lower compared to the previous model. Therefore, in this case, the $p_2$ parameter in the exponential term is allowing for a more flexible description of the Universe, and the data constraints favour a lower matter density to be consistent with the observed acceleration. The CC+$\mathrm{PN}^+\,\&\,$SH0ES data set combination yields the lowest value of $\Omega_{m,0}$, which is $\Omega_{m,0} = 0.269^{+0.046}_{-0.065}$. In tandem, the highest value for the Hubble constant is obtained for the same data set combination giving a value of $H_0 = 71.86^{+0.97}_{-0.99} \,{\rm km\, s}^{-1} {\rm Mpc}^{-1}$.

By design of the model itself, the parameter $\frac{1}{p_2}$ is positive throughout.  In comparison to the $f_1$CDM model, the parameter values of $f_2$CDM tend to fall within 2$\sigma$ of the $\Lambda$CDM limit instead of $1\sigma$. Therefore, the $f_2$CDM model is slightly further away from strongly supporting the $\Lambda$CDM model.

The inclusion of the $\mathrm{PN}^+\,\&\,$SH0ES dataset has a noticeable impact on the MCMC runs and the resulting model parameters. While the results are still in agreement with those obtained from the PN dataset alone, the uncertainties in the parameters, especially the Hubble constant, are significantly reduced. This makes the $\mathrm{PN}^+\,\&\,$SH0ES dataset useful for comparative purposes with $\Lambda$CDM. Further comparisons and statistical analyses with $\Lambda$CDM are discussed in Sec.~\ref{sec:model_comparison}.

\begin{table}
    \centering
    \small
    \caption{Results for the $f_2$CDM (Linder) model, where the first column lists the data sets used to constrain the parameters. The second to fourth columns display the constrained parameters, namely $H_0$, $\Omega_{m,0}$, and $\frac{1}{p_2}$, while the last column shows the nuisance parameter $M$.}
    \label{tab:LM}
    \begin{tabular}{ccccc}
        \hline
		Data sets & $H_0 \mathrm{\hspace{0.15cm}[km \hspace{0.1cm} s^{-1} \hspace{0.1cm}Mpc ^{-1}]}$ & $\Omega_{m,0}$ & $\frac{1}{p_{2}}$ & $M$ \\ 
		\hline
		CC + PN & $68.7^{+1.8}_{-1.7}$ & $0.298^{+0.031}_{-0.036}$ & $0.11^{+0.22}_{-0.11}$ & $-19.433^{+0.117}_{-0.083}$ \\ 
		CC + PN + BAO & $66.9^{+1.5}_{-1.6}$ & $0.294\pm 0.016$ & $0.22^{+0.12}_{-0.15}$ & $-19.38^{+0.22}_{-0.35}$ \\ 
		CC + $\mathrm{PN}^+\,\&\,$SH0ES & $71.86^{+0.97}_{-0.99}$ & $0.269^{+0.046}_{-0.065}$ & $0.39^{+0.29}_{-0.25}$ & $-19.287^{+0.048}_{-0.032}$ \\ 
		CC + $\mathrm{PN}^+\,\&\,$SH0ES + BAO & $70.79\pm 0.71$ & $0.328^{+0.013}_{-0.012}$ & $0.052^{+0.104}_{-0.038}$ & $-19.322^{+0.026}_{-0.033}$ \\ 
		\hline
    \end{tabular}
\end{table}

\subsection{Exponential Model}

The third model is motivated by works in $f(\lc{R})$ gravity \cite{Linder:2009jz}, in which an exponential model is again taken into consideration. Indeed, Nesseris et al. in Ref.\cite{Nesseris:2013jea}, propose a variant of the Linder model where the function $\mathcal{F}_3$ is given by an exponential function with two constants $\alpha_3$ and $p_3$ as parameters
\begin{equation}
    \mathcal{F}_3 = \alpha_3 T_0 \left(1- \mathrm{Exp}\left[-p_3 T/T_0\right] \right) \,.
\end{equation}
The constant $\alpha_3$ can be determined by evaluating the Friedmann equation at current times and is given by
\begin{equation}
    \alpha_3 = \frac{1-\Omega_{m,0} - \Omega_{r,0}}{(1+2p_3)e^{-p_3}-1}\,.
\end{equation}
The Friedmann equation for this model is therefore obtained using Eq.~\ref{eq:Friedmann_1} and substituting the above equations such that
\begin{equation}
    E^2\left(z\right) = \Omega_{m_0} \left(1+z\right)^3 + \Omega_{r_0}\left(1+z\right)^4 + \frac{1 - \Omega_{m_0} - \Omega_{r_0}}{(1 + 2p_3 )e^{-p_3} - 1} \left[\left(1 + 2p_3 E^2 (z)\right)\text{Exp}\left[-p_3 E^2 (z)\right] - 1\right]\,,
\end{equation}
The behavior of this model is similar to $f_2$CDM in the sense that as $p_3 \rightarrow \infty$, it tends towards $\Lambda$CDM. For numerical stability, the analysis is performed in terms of $1/p_3$ instead, such that the limit of $\Lambda$CDM corresponds to $1/p_3 \rightarrow 0^+$.

\begin{figure}
    \centering
    \includegraphics[width = \linewidth]{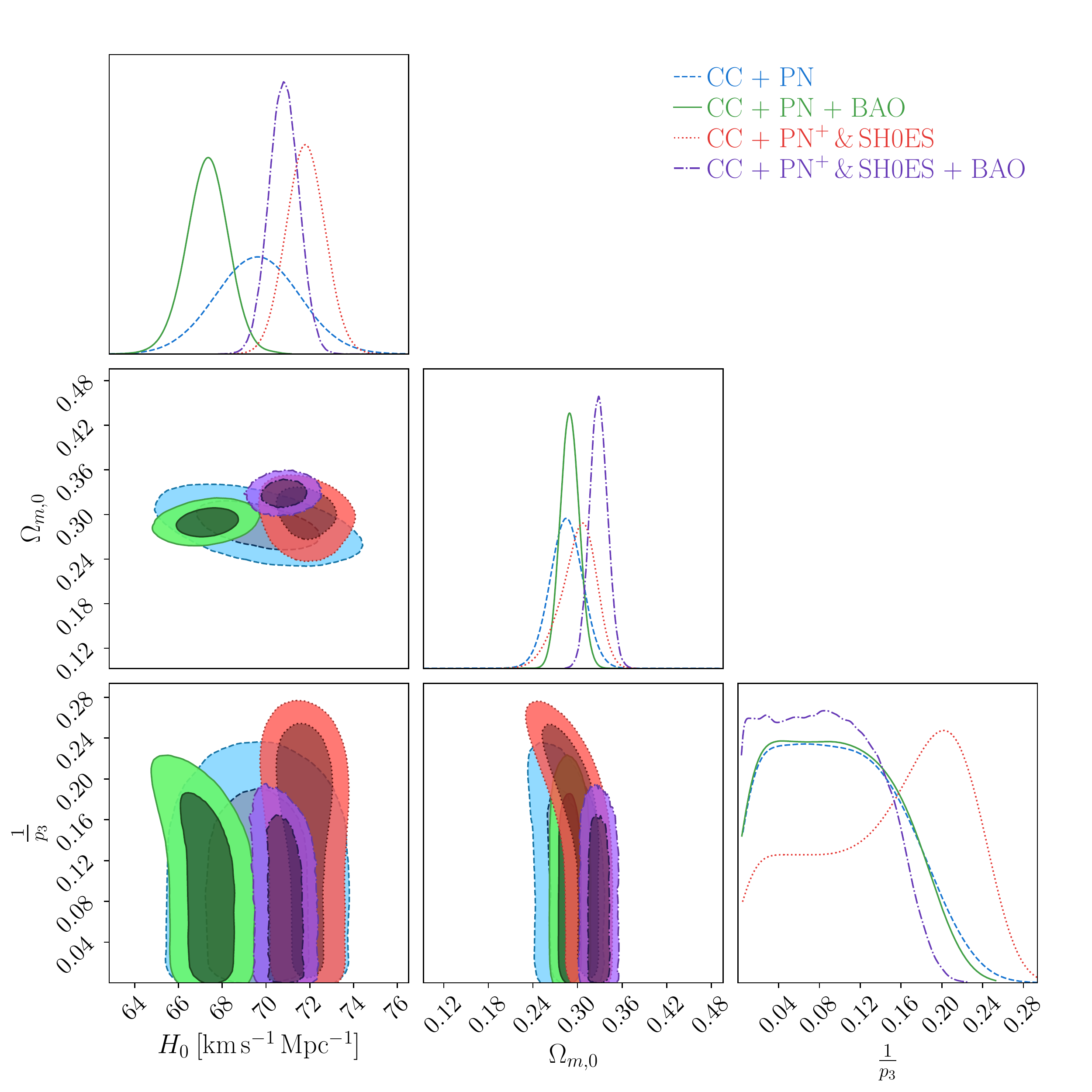}
    \caption{Confidence contours and posteriors for $f_3$CDM for the parameters $H_0$, $\Omega_{m,0}$ and $\frac{1}{p_3}$. The blue and green contours represent data set combinations that include PN data set, while the red and purple contours show combinations that also include the $\mathrm{PN}^+\,\&\,$SH0ES data sets.}
    \label{fig:LM2_CPB}
\end{figure}

The posterior and confidence levels for the $f_3$CDM model are presented in Fig.~\ref{fig:LM2_CPB}. Even though this model is a variant of the Linder model, the removal of the square root has had a significant impact on the constraints, in particular on the $\Omega_{m,0}$ parameter. Unlike the previous models, the degeneracy between $H_0$ and $\Omega_{m,0}$ parameters is no longer significant, but the correlation between $\Omega_{m,0}$ and $\frac{1}{p_3}$ is emphasized.

The constrained values for the parameters of the $f_3$CDM model are presented in Table~\ref{tab:LM2} which exhibit stricter and tighter confidence levels in the density parameter. Notably, the highest value of $H_0$ is once again obtained for the CC+$\mathrm{PN}^+\,\&\,$SH0ES data set combination, with $H_0 = 71.80 \pm 0.89,{\rm km\, s}^{-1} {\rm Mpc}^{-1}$. The value of $H_0$ obtained for CC+$\mathrm{PN}^+\,\&\,$SH0ES in the $f_3$CDM model is consistent with the previous corresponding values. However, the difference between the $H_0$ values for CC+$\mathrm{PN}^+\,\&\,$SH0ES and CC+$\mathrm{PN}^+\,\&\,$SH0ES +BAO is slightly larger than that obtained for $f_1$CDM. This implies that the value of $H_0$ for CC+$\mathrm{PN}^+\,\&\,$SH0ES+BAO is slightly lower in the $f_3$CDM model.

With regards to the $p_3$ parameter, the resulting values are closer to the $\Lambda$CDM limit when compared to the previous model. However, the uncertainties still suggest a deviation at the $2\sigma$ level from $\Lambda$CDM. These results obtained will be further analyzed and statistically compared with $\Lambda$CDM in the next section.

\begin{table}
    \centering
    \small
    \caption{Results for the $f_3$CDM model, where the first column lists the data sets used to constrain the parameters. The second to fourth columns display the constrained parameters, namely $H_0$, $\Omega_{m,0}$, and $\frac{1}{p_3}$, while the last column shows the nuisance parameter $M$.}
    \label{tab:LM2}
    \begin{tabular}{ccccc}
        \hline
		Data Sets & $H_0 \mathrm{\hspace{0.15cm}[km \hspace{0.1cm} s^{-1} \hspace{0.1cm}Mpc ^{-1}]}$ & $\Omega_{m,0}$ & $\frac{1}{p_{3}}$ & $M$ \\ 
		\hline
		CC + PN & $69.6^{+1.9}_{-2.0}$ & $0.286\pm 0.022$ & $0.065^{+0.082}_{-0.050}$ & $-19.367^{+0.054}_{-0.057}$ \\ 
		CC + PN + BAO & $67.35^{+0.94}_{-0.97}$ & $0.289\pm 0.013$ & $0.043^{+0.101}_{-0.026}$ & $-19.441^{+0.032}_{-0.031}$ \\ 
		CC + $\mathrm{PN}^+\,\&\,$SH0ES & $71.80\pm 0.89$ & $0.307^{+0.020}_{-0.026}$ & $0.201^{+0.045}_{-0.114}$ & $-19.302^{+0.033}_{-0.021}$ \\ 
		CC + $\mathrm{PN}^+\,\&\,$SH0ES + BAO & $70.80^{+0.70}_{-0.66}$ & $0.329\pm 0.012$ & $0.086^{+0.035}_{-0.081}$ & $-19.259\pm 0.077$ \\ 
		\hline
    \end{tabular}
\end{table}


\section{Model Comparison} \label{sec:model_comparison}

We evaluate the performance of each $f_i$CDM model and dataset by computing their respective minimum $\chi^2_\mathrm{min}$ values, obtained from the maximum likelihood $L_\mathrm{max}$ since $\chi^2_\mathrm{min} = -2 \ln L_\mathrm{max} $. Additionally, we compare the models against the standard $\Lambda$CDM by using the Akaike Information Criteria (AIC), which accounts for both the goodness of fit (measured by $\chi^2_\mathrm{min}$) and the complexity of the model (determined by the number of parameters $n$). The AIC is defined as 
\begin{equation}
    \mathrm{AIC} = \chi^2_\mathrm{min} + 2n \,.    
\end{equation}
In practice, a lower value of the AIC indicates that a model fits the data better, while also taking into account the complexity of the model. The AIC penalizes models that have more parameters, even if they provide a better fit to the data. This means that a model with a lower AIC is preferred over a model with a higher AIC, as long as the difference in AIC is significant enough.

In addition, we also examine the Bayesian Information Criterion (BIC), which is similar to AIC but it puts more weight on the complexity of the model than AIC does and is defined as
\begin{equation}
\mathrm{BIC} = \chi^2_\mathrm{min} + n \ln m \,,
\end{equation}
where $m$ is the sample size of the observational data combination. The BIC has the same goal as the AIC, that is, to balance the fit of the model to the data against the complexity of the model. However, the BIC tends to penalize models with more parameters more heavily than AIC does as it takes the logarithm of the sample size, so the penalty for more parameters becomes more severe as the sample size increases. In practical terms, comparing the BIC values of two models can help determine which one is more supported by the data, in which models with lower BIC values are favored as long as the difference is sufficiently large.

To compare the performance of various models using different combinations of data sets, we calculate the differences in AIC and BIC between each model and the reference model $\Lambda$CDM. The constrained parameters for $\Lambda$CDM model for each data set combination can be found in Table~\ref{tab:LCDM} in the Appendix~\ref{sec:LCDM}. Smaller values of $\Delta$AIC and $\Delta$BIC suggest that the model with the chosen data set is more similar to the $\Lambda$CDM model, indicating better performance. Indeed, Tables~\ref{tab:Compare_CCPN} and \ref{tab:Compare_CCPNBAO} provide the values for various statistical measures, such as $\chi^2_{\mathrm{min}}$, $\Delta$AIC $= \Delta \chi^2_\mathrm{min} + 2 \Delta n$, and $\Delta$BIC $= \Delta \chi^2_\mathrm{min} + \Delta n \ln m$, for each model.  Specifically, Table \ref{tab:Compare_CCPN} compares the models that use CC+PN with the ones that use CC+$\mathrm{PN}^+\,\&\,$SH0ES, whereas Table \ref{tab:Compare_CCPNBAO} compares the models that use CC+PN+BAO with the ones that use CC+$\mathrm{PN}^+\,\&\,$SH0ES+BAO. 

\begin{table}
\centering
\begin{tabular}{c||ccc|ccc}

    Model & \multicolumn{3}{c|}{CC + PN}& \multicolumn{3}{c}{CC+ $\mathrm{PN}^+\,\&\,$SH0ES}\\
    \hline
    \hline
     &$\chi^2_{\mathrm{min}}$ & $\Delta$AIC &  $\Delta$BIC & $\chi^2_{\mathrm{min}}$ & $\Delta$AIC &$\Delta$BIC \\
     \hline
    $\Lambda$CDM & 1041.49 & 0 &  0 & 1548.30 & 0 &  0           \\
    $f_1$CDM  & 1040.94 & 1.44 &   6.43       &  1546.64    & 0.34 &    5.80 \\
    $f_2$CDM   &    1041.49 &  2.00&      6.98     &   1546.67&     0.37  &  5.82         \\
    $f_3$CDM  &     1045.04& 5.54  &  10.53 & 1546.77    &   0.47  &    5.93      
\end{tabular}
    \caption{Results for each model that include $\chi^2_\mathrm{min}$, AIC, BIC, and their differences relative to the $\Lambda$CDM model (i.e., $\Delta$AIC and $\Delta$BIC). The left-hand side of the table presents the results obtained from the CC+PN data sets, while the right-hand side shows the results obtained from the CC+$\mathrm{PN}^+\,\&\,$SH0ES data sets.}
    \label{tab:Compare_CCPN}
\end{table}

\begin{table}
\centering
\begin{tabular}{c||ccc|ccc} 
    Model & \multicolumn{3}{c|}{CC+ PN + BAO}                    & \multicolumn{3}{c}{CC+ $\mathrm{PN}^+\,\&\,$SH0ES + BAO}                   \\
    \hline
    \hline
     &$\chi^2_{\mathrm{min}}$ & $\Delta$AIC & $\Delta$BIC & $\chi^2_{\mathrm{min}}$& $\Delta$AIC& $\Delta$BIC \\
     \hline
$\Lambda$CDM & 1057.46 & 0 & 0 & 1560.68 &  0 & 0           \\
$f_1$CDM  &   1057.13  &     1.68   & 6.68      &   1559.24 &     0.55 &  6.02     \\
$f_2$CDM   &    1056.52  &    1.06  &  6.06   &   1560.68&  1.99 &    7.46   \\
$f_3$CDM  &  1060.55 &    5.09 &  10.09  & 1560.68 &   1.99  &    7.47    
\end{tabular}
\caption{Results for each model that include $\chi^2_\mathrm{min}$, AIC, BIC, and their differences relative to the $\Lambda$CDM model (i.e., $\Delta$AIC and $\Delta$BIC). The left-hand side of the table presents the results obtained from the CC+PN+BAO data sets, while the right-hand side shows the results obtained from the CC+$\mathrm{PN}^+\,\&\,$SH0ES+BAO data sets.}
\label{tab:Compare_CCPNBAO}
\end{table}

Upon initial examination, it appears that the $\mathrm{PN}^+\,\&\,$SH0ES results in significantly lower values of $\Delta$AIC and $\Delta$BIC, despite the higher $\chi^2_\mathrm{min}$ value due to the increased number of data points. It is worth noting that the $\chi^2_\mathrm{min}$ values for the $f(T)$ models considered are slightly lower than that of the $\Lambda$CDM model for the CC+\,$\mathrm{PN}^+\,\&\,$SH0ES data set. Moreover, the values of $\Delta$AIC and $\Delta$BIC for the CC\,+\,$\mathrm{PN}^+\,\&\,$SH0ES are very close, indicating a stronger data set in which the constrained parameters are similar to those produced by the $\Lambda$CDM model. It seems that while CC\,+\,PN observations slightly support the $\Lambda$CDM model, the inclusion of $\mathrm{PN}^+\,\&\,$SH0ES data does not provide strong evidence in favor of the $\Lambda$CDM model over the considered $f(T)$ cosmological models given that both $\Delta$AIC and $\Delta$BIC are statistically comparable. Incorporating the BAO data set with the data sets reveals a similar trend, but to a lesser extent. However, for the $f_2$CDM model, the values for both $\Delta$AIC and $\Delta$BIC are higher for CC+$\mathrm{PN}^+\,\&\,$SH0ES+BAO, indicating that this model is not strongly supported by the observational data in comparison to the $\Lambda$CDM model.

\begin{figure}
    \centering
    \includegraphics[width = \textwidth,  trim={0 2cm 0 0},clip]{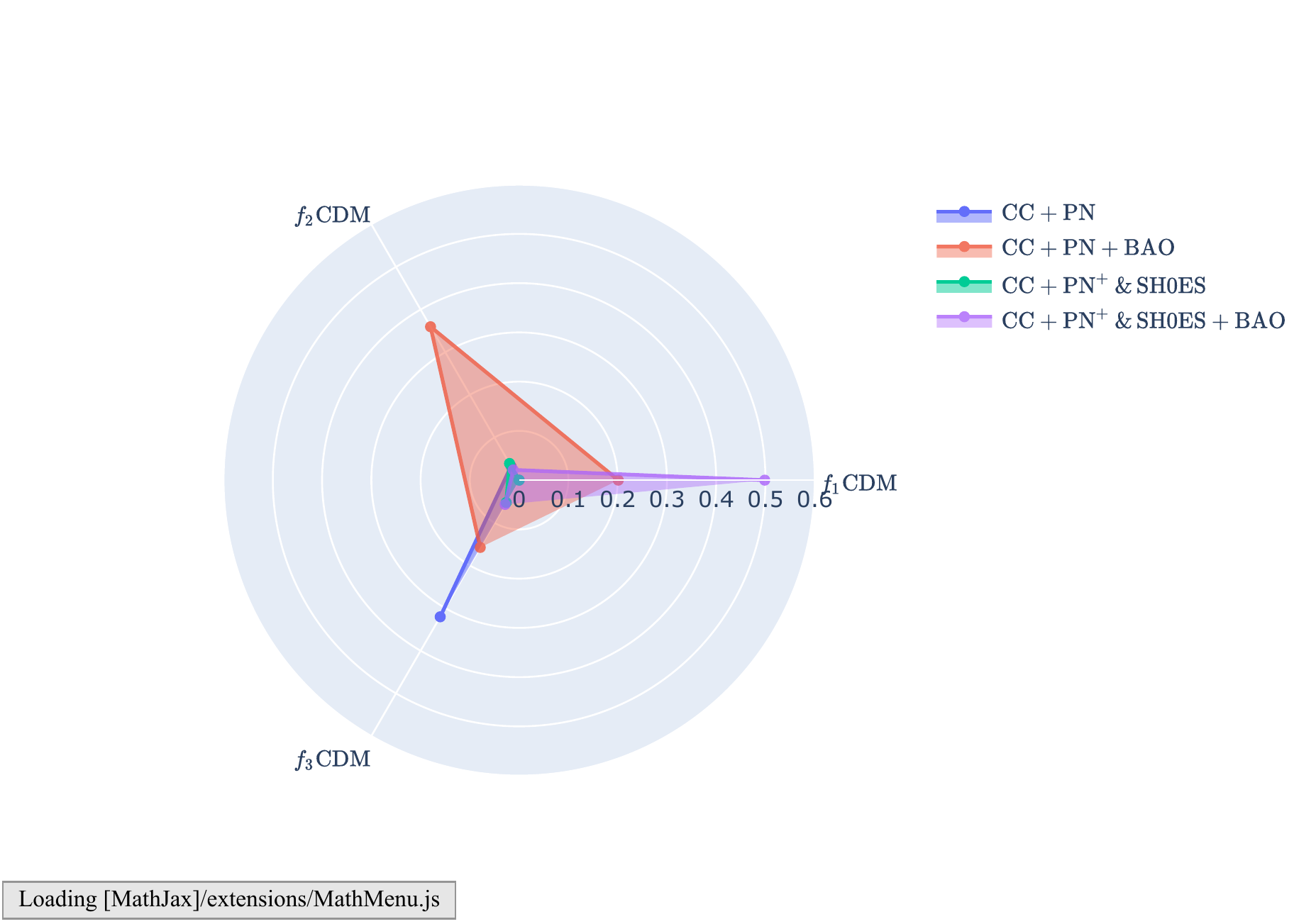}
    \caption{Distances, in units of standard deviations ($\sigma$), between the constrained values of $H_0$ and the $\Lambda$CDM value for different combinations of data sets, represented by different colors.
    }
    \label{fig:radar_plot_LCDM}
\end{figure}

\begin{figure}
    \centering
    \includegraphics[width= \textwidth, trim={0 1cm 0 0},clip]{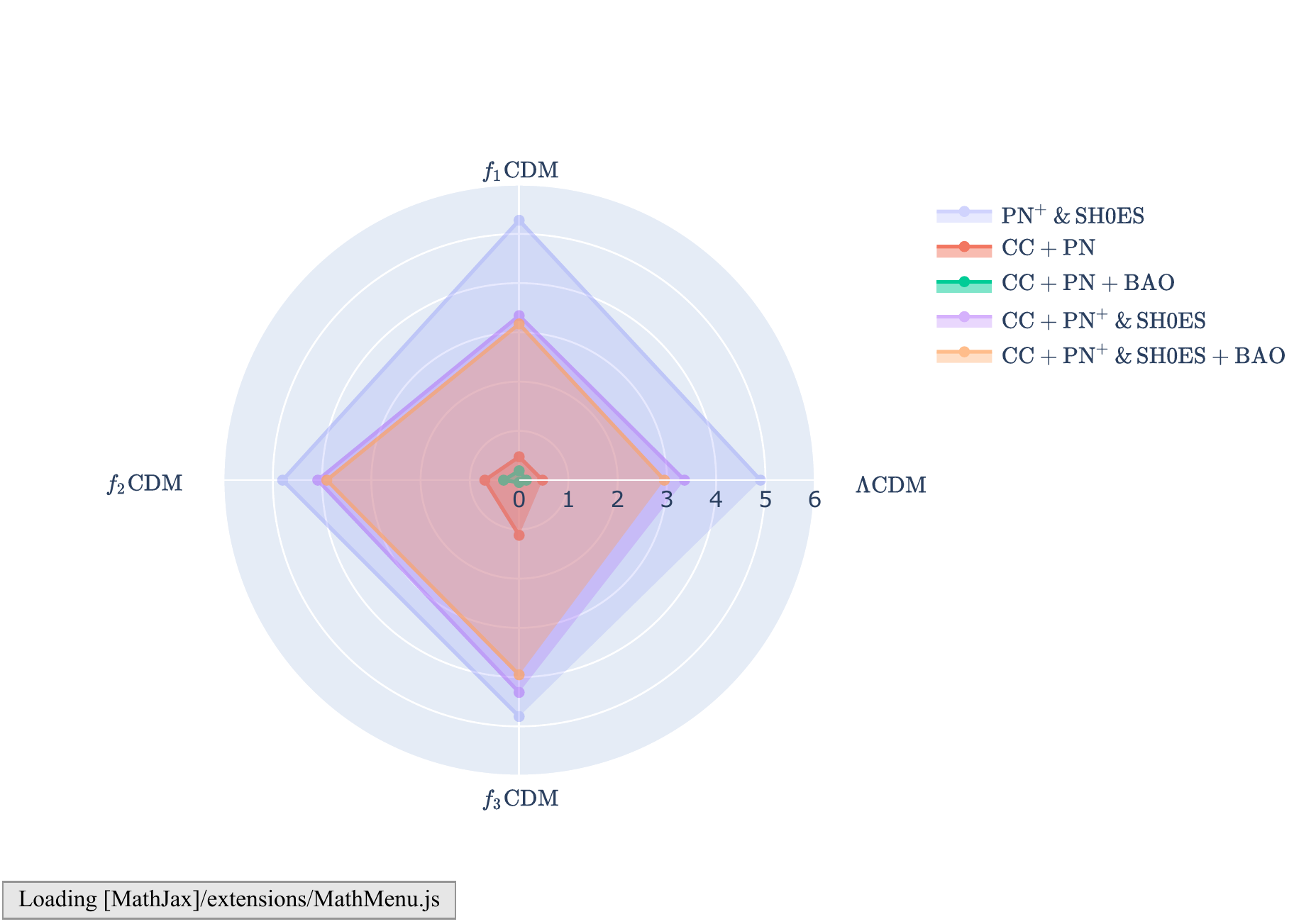}
    \caption{Distances, in units of standard deviations ($\sigma$), between the constrained values of $H_0$ for different combinations of data sets, represented by different colours and the P18 value.}
    \label{fig:Radar_plot_P18}
\end{figure}
The previous analysis is further supported by Fig.~\ref{fig:radar_plot_LCDM}, which compares the constrained $H_0$ values obtained from the $f(T)$ models to those obtained from the corresponding $\Lambda$CDM model. The figure shows that, for each data set combination represented by different colors, the $H_0$ values obtained from the $f(T)$ models are within $1 \sigma$ of the corresponding $\Lambda$CDM values. The plot provides a visualization of the variations in $H_0$ estimates across different data sets, with greater distances indicating larger discrepancies between the constrained and $\Lambda$CDM values of $H_0$. Therefore, the plot, suggests that the $H_0$ values obtained using the $f(T)$ models are comparable to those obtained using the $\Lambda$CDM model. 

In contrast, Fig.~\ref{fig:Radar_plot_P18} shows the difference in $\sigma$ units between the constrained $H_0$ values obtained from the MCMC analysis and the Planck 18 (P18) value of $H_0 = 67.4 \pm 0.5 {\rm km \,s\,}^{-1} {\rm Mpc}^{-1}$ \cite{Planck:2018vyg}. In this case, we also consider the $\mathrm{PN}^+\,\&\,$SH0ES data set on its own, for which the constrained $H_0$ values for each model are shown in Table~\ref{tab:PNp} in the Appendix~\ref{sec:PN+}. The plot clearly shows the $5 \sigma$ tension between the $\mathrm{PN}^+\,\&\,$SH0ES data set and the P18 value under the $\Lambda$CDM model. However, the inclusion of the CC data set at late-times appears to reduce the tension to around $3-4 \sigma$ for all models. Furthermore, inclusion of the BAO data set significantly reduces this tension, as expected, since the BAO data set captures the effects of the early Universe in agreement with the Planck CMB data set.

\begin{figure}
    \centering
    \includegraphics[width= \textwidth]{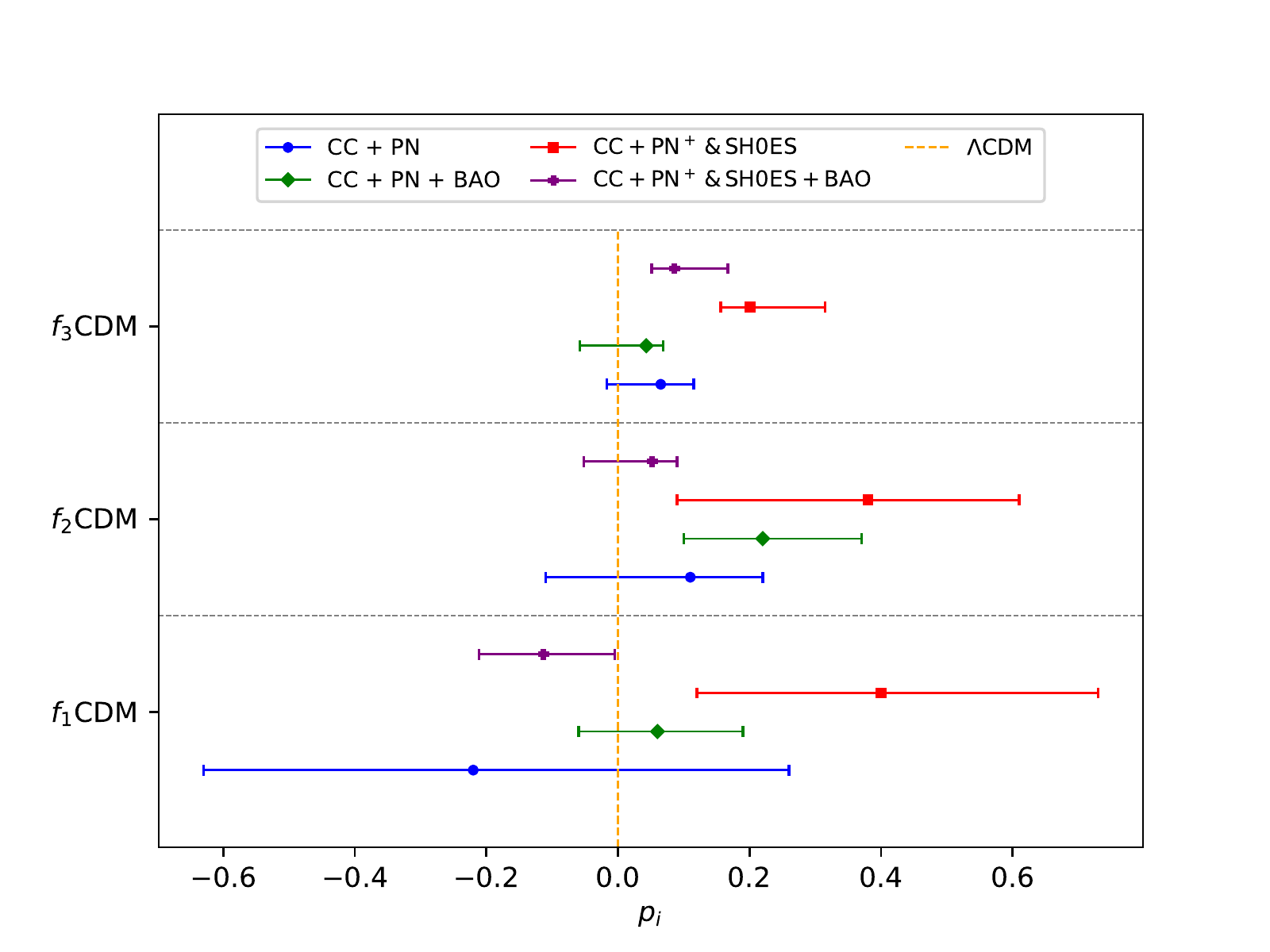}
    \caption{Values of the constrained model parameter $p_i$ which corresponds to $p_1$ for $f_1$CDM, $\frac{1}{p_2}$ for $f_2$CDM and $\frac{1}{p_3}$ for $f_3$CDM. Each colour represents a different data set combination while the orange line represents the $\Lambda$CDM value, i.e $p_i = 0$.}
    \label{fig:p_whisker_plot}
\end{figure}

Finally, we observe the effects that the $\mathrm{PN}^+\,\&\,$SH0ES has on the model parameter $p_i$, in the whisker plot Fig.~\ref{fig:p_whisker_plot}. The results indicate that the use of $\mathrm{PN}^+\,\&\,$SH0ES leads to a more tightly constrained estimate of $p_i$ compared to other methods, as previously observed. Notably, the CC+PN and CC+PN+BAO methods produce $p_i$ values that fall within 1$\sigma$ of the $\Lambda$CDM value. However, for CC+$\mathrm{PN}^+\,\&\,$SH0ES, this is not necessarily the case as the estimated $p_i$ values do not consistently fall within 1$\sigma$ of the $\Lambda$CDM value.

\section{Conclusion} \label{sec:conc}

In this work, we have presented a constraints analysis that examines the behaviour on the parameters of the $\mathrm{PN}^+\,\&\,$SH0ES over the PN data set. We evaluate three prominent models in $f(T)$ gravity and probe their performance against the two observational data sets by considering different data set combinations. Our primary objective was to compare the results obtained from the $\mathrm{PN}^+\,\&\,$SH0ES data sets to those of the PN catalog. We aimed to evaluate the differences in the outcomes of these data sets and assess their impact on the performance of the $f(T)$ gravity models under consideration. Indeed, for each model, we performed a full MCMC analysis obtaining observational constraints on the cosmological parameters for all different combinations of data. Additionally, we compared the performance of each model and data set to the standard model of cosmology using statistical indicators such as AIC and BIC. Finally, in light of the increasing tensions between cosmological observations, we have presented how the $H_0$ value compares the corresponding $\Lambda$CDM value and also with the P18 value.

We evaluated the performance of three models, namely $f_{1-3}$CDM, in which a continuous $\Lambda$CDM is present, and a specific setting of an additional model parameter recovers a constant cosmological constant contribution. For all models, the posterior and confidence contours immediately reveal the $\mathrm{PN}^+\,\&\,$SH0ES data set produced tighter constraints for the model parameters compared to the PN data set. Additionally, for all models considered, the $\mathrm{PN}^+\,\&\,$SH0ES data set produced higher values of $H_0$ due to its composition of the $\mathrm{PN}^+\,\&\,$SH0EScatalogue and the SH0ES Cepheid host distance anchors, which were consistent with previous SH0ES team results (R22). Notably, we obtained a consistent value of $H_0$ across all models for all different data set combinations. However, concerning the $\Omega_{m,0}$ parameter,$f_2$CDM and $f_3$CDM models, produce lower values than the $f_1$CDM model. The additional model parameter $p_i$, for the PN data set mostly fall within $1 \sigma$ of the $\Lambda$CDM model. However, with regards to $\mathrm{PN}^+\,\&\,$SH0ES they are mostly out of the $1 \sigma$ but within the $2 \sigma$ range. 

In Appendix~\ref{sec:LCDM}, we present the results obtained from the $\Lambda$CDM model, which we use for statistical comparisons. Our analysis revealed that the models under consideration are generally consistent with the $\Lambda$CDM model. Indeed, the statistical indicators, clearly indicate that the $\mathrm{PN}^+\,\&\,$SH0ES is a stronger data set as the constrained parameters are close to those produced by the $\Lambda$CDm model. In addition, the information criteria $\Delta$AIC and $\Delta$BIC suggest that the CC+PN data slightly support the $\Lambda$CDM model, whereas the $\mathrm{PN}^+\,\&\,$SH0ES data set does not provide strong evidence that supports the $\Lambda$CDM model over the $f(T)$ cosmological models, as indicated by their relatively small values.

Finally, incorporating the CC data with the $\mathrm{PN}^+\,\&\,$SH0ES data set reduces the $H_0$ tension to around $3\sigma$ (as illustrated in Fig.~\ref{fig:Radar_plot_P18}). Additionally, including the BAO data set also has an impact on the $H_0$ values, which are slightly reduced due to the effects from the early Universe. However, the contour plots in the triangular plots reveal an interesting point. When the BAO data set is included, the degeneracy between the parameters $H_0$ and $\Omega_{m,0}$ is broken, as demonstrated by the green and purple contours. Instead, a correlation between these parameters is revealed. Similarly, an anti-correlation between the $H_0$ and $p_i$ parameters is revealed when the BAO data set is included.
  
Therefore, our analysis provided insights into the behavior of the PN and the $\mathrm{PN}^+\,\&\,$SH0ES data sets and the performance of different models in $f(T)$ gravity. Our results suggest that the $\mathrm{PN}^+\,\&\,$SH0ES data set produces tighter constraints for model parameters and higher values of $H_0$ compared to the PN data set, and the inclusion of the CC and BAO data sets have a significant impact on the parameter degeneracies and tension in $H_0$. Overall, our analysis suggests that the $f(T)$ gravity models considered in this study provide a valuable framework for future investigations of modified gravity theories. We also intend to extend this work by considering CMB data frame from surveys such as the Planck Mission in order to be able to study the early phases of the Universe including analysis of the effects that such models would have on inflationary scenarios, for example.

\appendix
\section{\texorpdfstring{$\Lambda$}{lambda}CDM model} \label{sec:LCDM}
\begin{figure}
    \centering
    \includegraphics[width = \linewidth]{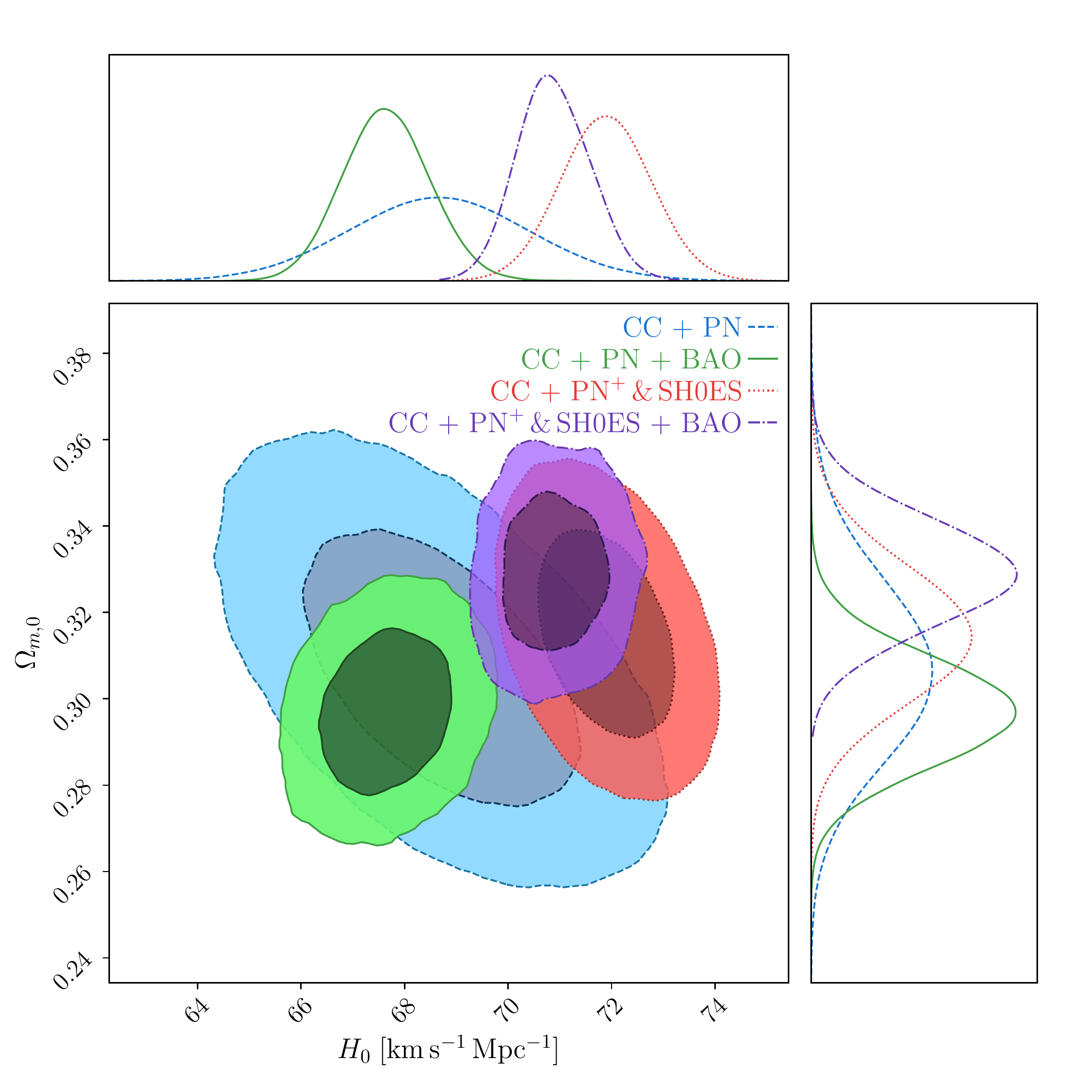}
    \caption{Confidence contours and posteriors for $\Lambda$CDM for the parameters $H_0$ and $\Omega_{m,0}$. The blue and green contours represent data set combinations that include PN data set, while the red and purple contours show combinations that also include the $\mathrm{PN}^+\,\&\,$SH0ES data sets.}
    \label{fig:LCDM}
\end{figure}

In Sec.~\ref{sec:model_comparison}, we provide comparisons between all $f_i$models and the respective $\Lambda$CDM MCMC runs. To this end, we provide here the results for $\Lambda$CDM. The plot in Fig.~\ref{fig:LCDM} display the posterior distributions ad confidence regions for the different combinations of data sets. The precise values of such runs are shown in Table~\ref{tab:LCDM}, in which as expected convergence for each data set combination occurs very fast giving nearly Gaussian uncertainties in each case.

\begin{table}
    \centering
    \caption{Results for the $\Lambda$CDM model, where the first column lists the data sets used to constrain the parameters. The second to fourth columns display the constrained parameters, namely $H_0$, $\Omega_{m,0}$, and the nuisance parameter $M$.}
    \label{tab:LCDM}
    \begin{tabular}{cccc}
        \hline
		Data Sets & $H_0 \mathrm{\hspace{0.15cm}[km \hspace{0.1cm} s^{-1} \hspace{0.1cm}Mpc ^{-1}]}$ & $\Omega_{m,0}$ & $M$ \\ 
		\hline
		CC + PN & $68.6^{+1.8}_{-1.7}$ & $0.306 \pm 0.021$ & $-19.383^{+0.050}_{-0.053}$ \\ 
		CC + PN + BAO & $67.59^{+0.89}_{-0.81}$ & $0.297\pm 0.013$ & $-19.419^{+0.026}_{-0.033}$ \\ 
		CC + $\mathrm{PN}^+\,\&\,$SH0ES & $71.88^{+0.88}_{-0.87}$ & $0.315\pm 0.016$ & $-19.298\pm 0.025$ \\ 
		CC + $\mathrm{PN}^+\,\&\,$SH0ES + BAO & $70.76^{+0.80}_{-0.64}$ & $0.329\pm 0.013$ & $-19.326^{+0.024}_{-0.022}$ \\ 
		\hline
    \end{tabular}
\end{table}

\section{\texorpdfstring{$\mathrm{PN}^+\,\&\,$}{}SH0ES parameter constraints} \label{sec:PN+}

To investigate the impact of the different data set combinations on the $H_0$ tension, we performed an MCMC analysis using only the $\mathrm{PN}^+\,\&\,$SH0ES data set as well. We then compared the deviation in units of $\sigma$ between the resulting $H_0$ values for each model and each data set combination with that of P18, as shown in Fig.~\ref{fig:Radar_plot_P18}. The constrained parameter values for each model obtained from this MCMC analysis are presented in Table~\ref{tab:PNp}.

\begin{table}
    \centering
    \caption{Results for the constrained parameters using the $\mathrm{PN}^+\,\&\,$SH0ES data set for each model considered in the analysis section.}
    \label{tab:PNp}
    \begin{tabular}{ccccc}
        \hline
		Model & $H_0 \mathrm{\hspace{0.15cm}[km \hspace{0.1cm} s^{-1} \hspace{0.1cm}Mpc ^{-1}]}$ & $\Omega_{m,0}$ & $p_i$ & $M$ \\ 
        \hline
		$\Lambda$CDM & $73.4 \pm 1.1$ & $0.334^{+0.021}_{-0.020}$ & -- &$-19.247 \pm 0.033$ \\ 
        $f_1$CDM & $73.3\pm 1.0$ & $0.331^{+0.044}_{-0.070}$ & $0.28^{+0.22}_{-0.37}$ & $-19.248^{+0.030}_{-0.029}$ \\ 
        $f_2$CDM & $73.2^{+1.1}_{-1.0}$ & $0.318^{+0.023}_{-0.102}$ & $0.33^{+0.32}_{-0.26}$ & $-19.259^{+0.044}_{-0.021}$ \\ 
        $f_3$CDM & $73.2\pm 1.1$ & $0.308^{+0.032}_{-0.099}$ & $0.33^{+0.34}_{-0.24}$ & $-19.225^{+0.040}_{-0.085}$ \\ 
        \hline
    \end{tabular}
\end{table}


\section*{Acknowledgements}\label{sec:acknowledgements}
The authors would like to acknowledge support from the Malta Digital Innovation Authority through the IntelliVerse grant. This paper is based upon work from COST Action CA21136 {\it Addressing observational tensions in cosmology with systematics and fundamental physics} (CosmoVerse) supported by COST (European Cooperation in Science and Technology). This research has been carried out using computational facilities procured through the European Regional Development Fund, Project No. ERDF-080 ``A supercomputing laboratory for the University of Malta''. This paper is based upon work from COST Action CA21136 {\it Addressing observational tensions in cosmology with systematics and fundamental physics} (CosmoVerse) supported by COST (European Cooperation in Science and Technology).
CE-R acknowledges the Royal Astronomical Society as FRAS 10147 and is supported by PAPIIT UNAM Project TA100122.


\bibliographystyle{JHEP}
\bibliography{references}

\providecommand{\href}[2]{#2}\begingroup\raggedright\begin{thebibliography}{10}

\bibitem{Misner:1974qy}
C.W.~Misner, K.S.~Thorne and J.A.~Wheeler, \emph{{Gravitation}}, W. H. Freeman,
  San Francisco (1973).

\bibitem{Clifton:2011jh}
T.~Clifton, P.G.~Ferreira, A.~Padilla and C.~Skordis, \emph{{Modified Gravity
  and Cosmology}},
  \href{https://doi.org/10.1016/j.physrep.2012.01.001}{\emph{Phys. Rept.}
  {\bfseries 513} (2012) 1} [\href{https://arxiv.org/abs/1106.2476}{{\ttfamily
  1106.2476}}].

\bibitem{Baudis:2016qwx}
L.~Baudis, \emph{{Dark matter detection}},
  \href{https://doi.org/10.1088/0954-3899/43/4/044001}{\emph{J. Phys. G}
  {\bfseries 43} (2016) 044001}.

\bibitem{Bertone:2004pz}
G.~Bertone, D.~Hooper and J.~Silk, \emph{{Particle dark matter: Evidence,
  candidates and constraints}},
  \href{https://doi.org/10.1016/j.physrep.2004.08.031}{\emph{Phys. Rept.}
  {\bfseries 405} (2005) 279}
  [\href{https://arxiv.org/abs/hep-ph/0404175}{{\ttfamily hep-ph/0404175}}].

\bibitem{Peebles:2002gy}
P.J.E.~Peebles and B.~Ratra, \emph{{The Cosmological Constant and Dark
  Energy}}, \href{https://doi.org/10.1103/RevModPhys.75.559}{\emph{Rev. Mod.
  Phys.} {\bfseries 75} (2003) 559}
  [\href{https://arxiv.org/abs/astro-ph/0207347}{{\ttfamily
  astro-ph/0207347}}].

\bibitem{Copeland:2006wr}
E.J.~Copeland, M.~Sami and S.~Tsujikawa, \emph{{Dynamics of dark energy}},
  \href{https://doi.org/10.1142/S021827180600942X}{\emph{Int. J. Mod. Phys. D}
  {\bfseries 15} (2006) 1753}
  [\href{https://arxiv.org/abs/hep-th/0603057}{{\ttfamily hep-th/0603057}}].

\bibitem{Weinberg:1988cp}
S.~Weinberg, \emph{{The Cosmological Constant Problem}},
  \href{https://doi.org/10.1103/RevModPhys.61.1}{\emph{Rev. Mod. Phys.}
  {\bfseries 61} (1989) 1}.

\bibitem{Gaitskell:2004gd}
R.J.~Gaitskell, \emph{{Direct detection of dark matter}},
  \href{https://doi.org/10.1146/annurev.nucl.54.070103.181244}{\emph{Ann. Rev.
  Nucl. Part. Sci.} {\bfseries 54} (2004) 315}.

\bibitem{DiValentino:2020zio}
E.~Di~Valentino et~al., \emph{{Snowmass2021 - Letter of interest cosmology
  intertwined II: The hubble constant tension}},
  \href{https://doi.org/10.1016/j.astropartphys.2021.102605}{\emph{Astropart.
  Phys.} {\bfseries 131} (2021) 102605}
  [\href{https://arxiv.org/abs/2008.11284}{{\ttfamily 2008.11284}}].

\bibitem{Riess:2019cxk}
A.G.~Riess, S.~Casertano, W.~Yuan, L.M.~Macri and D.~Scolnic, \emph{{Large
  Magellanic Cloud Cepheid Standards Provide a 1\% Foundation for the
  Determination of the Hubble Constant and Stronger Evidence for Physics beyond
  $\Lambda$CDM}},
  \href{https://doi.org/10.3847/1538-4357/ab1422}{\emph{Astrophys. J.}
  {\bfseries 876} (2019) 85}
  [\href{https://arxiv.org/abs/1903.07603}{{\ttfamily 1903.07603}}].

\bibitem{Wong:2019kwg}
K.C.~Wong et~al., \emph{{H0LiCOW XIII. A 2.4\% measurement of $H_{0}$ from
  lensed quasars: $5.3\sigma$ tension between early and late-Universe probes}},
   \href{https://arxiv.org/abs/1907.04869}{{\ttfamily 1907.04869}}.

\bibitem{Aghanim:2018eyx}
{\scshape Planck} collaboration, \emph{{Planck 2018 results. VI. Cosmological
  parameters}},  \href{https://arxiv.org/abs/1807.06209}{{\ttfamily
  1807.06209}}.

\bibitem{Ade:2015xua}
{\scshape Planck} collaboration, \emph{{Planck 2015 results. XIII. Cosmological
  parameters}},
  \href{https://doi.org/10.1051/0004-6361/201525830}{\emph{Astron. Astrophys.}
  {\bfseries 594} (2016) A13}
  [\href{https://arxiv.org/abs/1502.01589}{{\ttfamily 1502.01589}}].

\bibitem{Riess:2020sih}
A.G.~Riess, \emph{{The Expansion of the Universe is Faster than Expected}},
  \href{https://doi.org/10.1038/s42254-019-0137-0}{\emph{Nature Rev. Phys.}
  {\bfseries 2} (2019) 10} [\href{https://arxiv.org/abs/2001.03624}{{\ttfamily
  2001.03624}}].

\bibitem{Pesce:2020xfe}
D.W.~Pesce et~al., \emph{{The Megamaser Cosmology Project. XIII. Combined
  Hubble constant constraints}},
  \href{https://doi.org/10.3847/2041-8213/ab75f0}{\emph{Astrophys. J. Lett.}
  {\bfseries 891} (2020) L1}
  [\href{https://arxiv.org/abs/2001.09213}{{\ttfamily 2001.09213}}].

\bibitem{deJaeger:2020zpb}
T.~de~Jaeger, B.E.~Stahl, W.~Zheng, A.V.~Filippenko, A.G.~Riess and L.~Galbany,
  \emph{{A measurement of the Hubble constant from Type II supernovae}},
  \href{https://doi.org/10.1093/mnras/staa1801}{\emph{Mon. Not. Roy. Astron.
  Soc.} {\bfseries 496} (2020) 3402}
  [\href{https://arxiv.org/abs/2006.03412}{{\ttfamily 2006.03412}}].

\bibitem{Baker:2019nia}
J.~Baker et~al., \emph{{The Laser Interferometer Space Antenna: Unveiling the
  Millihertz Gravitational Wave Sky}},
  \href{https://arxiv.org/abs/1907.06482}{{\ttfamily 1907.06482}}.

\bibitem{2017arXiv170200786A}
P.~{Amaro-Seoane}, H.~{Audley} et~al., \emph{{Laser Interferometer Space
  Antenna}}, {\emph{arXiv e-prints} (2017) arXiv:1702.00786}
  [\href{https://arxiv.org/abs/1702.00786}{{\ttfamily 1702.00786}}].

\bibitem{Barack:2018yly}
L.~Barack et~al., \emph{{Black holes, gravitational waves and fundamental
  physics: a roadmap}},
  \href{https://doi.org/10.1088/1361-6382/ab0587}{\emph{Class. Quant. Grav.}
  {\bfseries 36} (2019) 143001}
  [\href{https://arxiv.org/abs/1806.05195}{{\ttfamily 1806.05195}}].

\bibitem{Bernal:2016gxb}
J.L.~Bernal, L.~Verde and A.G.~Riess, \emph{{The trouble with $H_0$}},
  \href{https://doi.org/10.1088/1475-7516/2016/10/019}{\emph{JCAP} {\bfseries
  10} (2016) 019} [\href{https://arxiv.org/abs/1607.05617}{{\ttfamily
  1607.05617}}].

\bibitem{DiValentino:2021izs}
E.~Di~Valentino, O.~Mena, S.~Pan, L.~Visinelli, W.~Yang, A.~Melchiorri et~al.,
  \emph{{In the Realm of the Hubble tension $-$ a Review of Solutions}},
  \href{https://arxiv.org/abs/2103.01183}{{\ttfamily 2103.01183}}.

\bibitem{Sotiriou:2008rp}
T.P.~Sotiriou and V.~Faraoni, \emph{{f(R) Theories Of Gravity}},
  \href{https://doi.org/10.1103/RevModPhys.82.451}{\emph{Rev. Mod. Phys.}
  {\bfseries 82} (2010) 451} [\href{https://arxiv.org/abs/0805.1726}{{\ttfamily
  0805.1726}}].

\bibitem{CANTATA:2021ktz}
{\scshape CANTATA} collaboration, \emph{{Modified Gravity and Cosmology: An
  Update by the CANTATA Network}},
  \href{https://arxiv.org/abs/2105.12582}{{\ttfamily 2105.12582}}.

\bibitem{misner1973gravitation}
C.~Misner, K.~Thorne and J.~Wheeler, \emph{Gravitation}, no.~pt. 3 in
  Gravitation, W. H. Freeman (1973).

\bibitem{nakahara2003geometry}
M.~Nakahara, \emph{Geometry, Topology and Physics, Second Edition}, Graduate
  student series in physics, Taylor \& Francis (2003).

\bibitem{Bahamonde:2021gfp}
S.~Bahamonde, K.F.~Dialektopoulos, C.~Escamilla-Rivera, G.~Farrugia, V.~Gakis,
  M.~Hendry et~al., \emph{{Teleparallel Gravity: From Theory to Cosmology}},
  \href{https://arxiv.org/abs/2106.13793}{{\ttfamily 2106.13793}}.

\bibitem{Aldrovandi:2013wha}
R.~Aldrovandi and J.G.~Pereira, \emph{{Teleparallel Gravity}: {An
  Introduction}}, Springer (2013),
  \href{https://doi.org/10.1007/978-94-007-5143-9}{10.1007/978-94-007-5143-9}.

\bibitem{Cai:2015emx}
Y.-F.~Cai, S.~Capozziello, M.~De~Laurentis and E.N.~Saridakis, \emph{{$f(T)$
  teleparallel gravity and cosmology}},
  \href{https://doi.org/10.1088/0034-4885/79/10/106901}{\emph{Rept. Prog.
  Phys.} {\bfseries 79} (2016) 106901}
  [\href{https://arxiv.org/abs/1511.07586}{{\ttfamily 1511.07586}}].

\bibitem{Krssak:2018ywd}
M.~Krssak, R.J.~van~den Hoogen, J.G.~Pereira, C.G.~B\"ohmer and A.A.~Coley,
  \emph{{Teleparallel theories of gravity: illuminating a fully invariant
  approach}}, \href{https://doi.org/10.1088/1361-6382/ab2e1f}{\emph{Class.
  Quant. Grav.} {\bfseries 36} (2019) 183001}
  [\href{https://arxiv.org/abs/1810.12932}{{\ttfamily 1810.12932}}].

\bibitem{Ferraro:2006jd}
R.~Ferraro and F.~Fiorini, \emph{{Modified teleparallel gravity: Inflation
  without inflaton}},
  \href{https://doi.org/10.1103/PhysRevD.75.084031}{\emph{Phys. Rev.}
  {\bfseries D75} (2007) 084031}
  [\href{https://arxiv.org/abs/gr-qc/0610067}{{\ttfamily gr-qc/0610067}}].

\bibitem{Ferraro:2008ey}
R.~Ferraro and F.~Fiorini, \emph{{On Born-Infeld Gravity in Weitzenbock
  spacetime}}, \href{https://doi.org/10.1103/PhysRevD.78.124019}{\emph{Phys.
  Rev.} {\bfseries D78} (2008) 124019}
  [\href{https://arxiv.org/abs/0812.1981}{{\ttfamily 0812.1981}}].

\bibitem{Bengochea:2008gz}
G.R.~Bengochea and R.~Ferraro, \emph{{Dark torsion as the cosmic speed-up}},
  \href{https://doi.org/10.1103/PhysRevD.79.124019}{\emph{Phys. Rev.}
  {\bfseries D79} (2009) 124019}
  [\href{https://arxiv.org/abs/0812.1205}{{\ttfamily 0812.1205}}].

\bibitem{Linder:2010py}
E.V.~Linder, \emph{{Einstein's Other Gravity and the Acceleration of the
  Universe}}, \href{https://doi.org/10.1103/PhysRevD.81.127301,
  10.1103/PhysRevD.82.109902}{\emph{Phys. Rev.} {\bfseries D81} (2010) 127301}
  [\href{https://arxiv.org/abs/1005.3039}{{\ttfamily 1005.3039}}].

\bibitem{Chen:2010va}
S.-H.~Chen, J.B.~Dent, S.~Dutta and E.N.~Saridakis, \emph{{Cosmological
  perturbations in f(T) gravity}},
  \href{https://doi.org/10.1103/PhysRevD.83.023508}{\emph{Phys. Rev.}
  {\bfseries D83} (2011) 023508}
  [\href{https://arxiv.org/abs/1008.1250}{{\ttfamily 1008.1250}}].

\bibitem{Bahamonde:2019zea}
S.~Bahamonde, K.~Flathmann and C.~Pfeifer, \emph{{Photon sphere and perihelion
  shift in weak $f(T)$ gravity}},
  \href{https://doi.org/10.1103/PhysRevD.100.084064}{\emph{Phys. Rev. D}
  {\bfseries 100} (2019) 084064}
  [\href{https://arxiv.org/abs/1907.10858}{{\ttfamily 1907.10858}}].

\bibitem{Paliathanasis:2017htk}
A.~Paliathanasis, J.~Levi~Said and J.D.~Barrow, \emph{{Stability of the Kasner
  Universe in f(T) Gravity}},
  \href{https://doi.org/10.1103/PhysRevD.97.044008}{\emph{Phys. Rev. D}
  {\bfseries 97} (2018) 044008}
  [\href{https://arxiv.org/abs/1709.03432}{{\ttfamily 1709.03432}}].

\bibitem{Farrugia:2020fcu}
G.~Farrugia, J.~Levi~Said and A.~Finch, \emph{{Gravitoelectromagnetism, Solar
  System Tests, and Weak-Field Solutions in $f (T,B)$ Gravity with
  Observational Constraints}},
  \href{https://doi.org/10.3390/universe6020034}{\emph{Universe} {\bfseries 6}
  (2020) 34} [\href{https://arxiv.org/abs/2002.08183}{{\ttfamily 2002.08183}}].

\bibitem{Bahamonde:2021srr}
S.~Bahamonde, A.~Golovnev, M.-J.~Guzm\'an, J.L.~Said and C.~Pfeifer,
  \emph{{Black holes in f(T,B) gravity: exact and perturbed solutions}},
  \href{https://doi.org/10.1088/1475-7516/2022/01/037}{\emph{JCAP} {\bfseries
  01} (2022) 037} [\href{https://arxiv.org/abs/2110.04087}{{\ttfamily
  2110.04087}}].

\bibitem{Bahamonde:2020bbc}
S.~Bahamonde, J.~Levi~Said and M.~Zubair, \emph{{Solar system tests in modified
  teleparallel gravity}},
  \href{https://doi.org/10.1088/1475-7516/2020/10/024}{\emph{JCAP} {\bfseries
  10} (2020) 024} [\href{https://arxiv.org/abs/2006.06750}{{\ttfamily
  2006.06750}}].

\bibitem{Farrugia:2016qqe}
G.~Farrugia and J.~Levi~Said, \emph{{Stability of the flat FLRW metric in
  $f(T)$ gravity}},
  \href{https://doi.org/10.1103/PhysRevD.94.124054}{\emph{Phys. Rev. D}
  {\bfseries 94} (2016) 124054}
  [\href{https://arxiv.org/abs/1701.00134}{{\ttfamily 1701.00134}}].

\bibitem{Finch:2018gkh}
A.~Finch and J.L.~Said, \emph{{Galactic Rotation Dynamics in f(T) gravity}},
  \href{https://doi.org/10.1140/epjc/s10052-018-6028-1}{\emph{Eur. Phys. J. C}
  {\bfseries 78} (2018) 560}
  [\href{https://arxiv.org/abs/1806.09677}{{\ttfamily 1806.09677}}].

\bibitem{Farrugia:2016xcw}
G.~Farrugia, J.~Levi~Said and M.L.~Ruggiero, \emph{{Solar System tests in f(T)
  gravity}}, \href{https://doi.org/10.1103/PhysRevD.93.104034}{\emph{Phys. Rev.
  D} {\bfseries 93} (2016) 104034}
  [\href{https://arxiv.org/abs/1605.07614}{{\ttfamily 1605.07614}}].

\bibitem{Iorio:2012cm}
L.~Iorio and E.N.~Saridakis, \emph{{Solar system constraints on f(T) gravity}},
  \href{https://doi.org/10.1111/j.1365-2966.2012.21995.x}{\emph{Mon. Not. Roy.
  Astron. Soc.} {\bfseries 427} (2012) 1555}
  [\href{https://arxiv.org/abs/1203.5781}{{\ttfamily 1203.5781}}].

\bibitem{Deng:2018ncg}
X.-M.~Deng, \emph{{Probing f(T) gravity with gravitational time advancement}},
  \href{https://doi.org/10.1088/1361-6382/aad391}{\emph{Class. Quant. Grav.}
  {\bfseries 35} (2018) 175013}.

\bibitem{Nesseris:2013jea}
S.~Nesseris, S.~Basilakos, E.N.~Saridakis and L.~Perivolaropoulos,
  \emph{{Viable $f(T)$ models are practically indistinguishable from
  $\Lambda$CDM}}, \href{https://doi.org/10.1103/PhysRevD.88.103010}{\emph{Phys.
  Rev. D} {\bfseries 88} (2013) 103010}
  [\href{https://arxiv.org/abs/1308.6142}{{\ttfamily 1308.6142}}].

\bibitem{Anagnostopoulos:2019miu}
F.K.~Anagnostopoulos, S.~Basilakos and E.N.~Saridakis, \emph{{Bayesian analysis
  of $f(T)$ gravity using $f\sigma_8$ data}},
  \href{https://doi.org/10.1103/PhysRevD.100.083517}{\emph{Phys. Rev. D}
  {\bfseries 100} (2019) 083517}
  [\href{https://arxiv.org/abs/1907.07533}{{\ttfamily 1907.07533}}].

\bibitem{Nunes:2018evm}
R.C.~Nunes, S.~Pan and E.N.~Saridakis, \emph{{New observational constraints on
  $f(T)$ gravity through gravitational-wave astronomy}},
  \href{https://doi.org/10.1103/PhysRevD.98.104055}{\emph{Phys. Rev. D}
  {\bfseries 98} (2018) 104055}
  [\href{https://arxiv.org/abs/1810.03942}{{\ttfamily 1810.03942}}].

\bibitem{Benetti:2020hxp}
M.~Benetti, S.~Capozziello and G.~Lambiase, \emph{{Updating constraints on f(T)
  teleparallel cosmology and the consistency with Big Bang Nucleosynthesis}},
  \href{https://doi.org/10.1093/mnras/staa3368}{\emph{Mon. Not. Roy. Astron.
  Soc.} {\bfseries 500} (2020) 1795}
  [\href{https://arxiv.org/abs/2006.15335}{{\ttfamily 2006.15335}}].

\bibitem{Freedman:2019jwv}
W.L.~Freedman et~al., \emph{{The Carnegie-Chicago Hubble Program. VIII. An
  independent determination of the Hubble constant based on the Tip of the Red
  Giant Branch}},
  \href{https://doi.org/10.3847/1538-4357/ab2f73}{\emph{Astrophys. J.}
  {\bfseries 882} (2019) 34}
  [\href{https://arxiv.org/abs/1907.05922}{{\ttfamily 1907.05922}}].

\bibitem{Briffa:2021nxg}
R.~Briffa, C.~Escamilla-Rivera, J.~Said~Levi, J.~Mifsud and N.L.~Pullicino,
  \emph{{Impact of $H_0$ priors on $f(T)$ late time cosmology}},
  \href{https://doi.org/10.1140/epjp/s13360-022-02725-4}{\emph{Eur. Phys. J.
  Plus} {\bfseries 137} (2022) 532}
  [\href{https://arxiv.org/abs/2108.03853}{{\ttfamily 2108.03853}}].

\bibitem{Scolnic:2017caz}
D.M.~Scolnic et~al., \emph{{The complete light-curve sample of
  spectroscopically confirmed SNe Ia from Pan-STARRS1 and cosmological
  constraints from the combined Pantheon Sample}},
  \href{https://doi.org/10.3847/1538-4357/aab9bb}{\emph{Astrophys. J.}
  {\bfseries 859} (2018) 101}
  [\href{https://arxiv.org/abs/1710.00845}{{\ttfamily 1710.00845}}].

\bibitem{Brout:2021mpj}
D.~Brout et~al., \emph{{The Pantheon+ Analysis: SuperCal-fragilistic Cross
  Calibration, Retrained SALT2 Light-curve Model, and Calibration Systematic
  Uncertainty}},
  \href{https://doi.org/10.3847/1538-4357/ac8bcc}{\emph{Astrophys. J.}
  {\bfseries 938} (2022) 111}
  [\href{https://arxiv.org/abs/2112.03864}{{\ttfamily 2112.03864}}].

\bibitem{Riess:2021jrx}
A.G.~Riess et~al., \emph{{A Comprehensive Measurement of the Local Value of the
  Hubble Constant with 1 km s$^{-1}$ Mpc$^{-1}$ Uncertainty from the Hubble
  Space Telescope and the SH0ES Team}},
  \href{https://doi.org/10.3847/2041-8213/ac5c5b}{\emph{Astrophys. J. Lett.}
  {\bfseries 934} (2022) L7}
  [\href{https://arxiv.org/abs/2112.04510}{{\ttfamily 2112.04510}}].

\bibitem{Scolnic:2021amr}
D.~Scolnic et~al., \emph{{The Pantheon+ Analysis: The Full Data Set and
  Light-curve Release}},
  \href{https://doi.org/10.3847/1538-4357/ac8b7a}{\emph{Astrophys. J.}
  {\bfseries 938} (2022) 113}
  [\href{https://arxiv.org/abs/2112.03863}{{\ttfamily 2112.03863}}].

\bibitem{Hayashi:1979qx}
K.~Hayashi and T.~Shirafuji, \emph{{New General Relativity}},
  \href{https://doi.org/10.1103/PhysRevD.19.3524}{\emph{Phys. Rev. D}
  {\bfseries 19} (1979) 3524}.

\bibitem{Weitzenbock1923}
R.~Weitzenb\"{o}ock, \emph{`Invariantentheorie'}, Noordhoff, Gronningen (1923).

\bibitem{Bahamonde:2015zma}
S.~Bahamonde, C.G.~B\"ohmer and M.~Wright, \emph{{Modified teleparallel
  theories of gravity}},
  \href{https://doi.org/10.1103/PhysRevD.92.104042}{\emph{Phys. Rev. D}
  {\bfseries 92} (2015) 104042}
  [\href{https://arxiv.org/abs/1508.05120}{{\ttfamily 1508.05120}}].

\bibitem{RezaeiAkbarieh:2018ijw}
A.~Rezaei~Akbarieh and Y.~Izadi, \emph{{Tachyon Inflation in Teleparallel
  Gravity}}, \href{https://doi.org/10.1140/epjc/s10052-019-6819-z}{\emph{Eur.
  Phys. J. C} {\bfseries 79} (2019) 366}
  [\href{https://arxiv.org/abs/1812.06649}{{\ttfamily 1812.06649}}].

\bibitem{Krssak:2015oua}
M.~Kr\v{s}\v{s}\'ak and E.N.~Saridakis, \emph{{The covariant formulation of
  f(T) gravity}},
  \href{https://doi.org/10.1088/0264-9381/33/11/115009}{\emph{Class. Quant.
  Grav.} {\bfseries 33} (2016) 115009}
  [\href{https://arxiv.org/abs/1510.08432}{{\ttfamily 1510.08432}}].

\bibitem{Tamanini:2012hg}
N.~Tamanini and C.G.~Boehmer, \emph{{Good and bad tetrads in f(T) gravity}},
  \href{https://doi.org/10.1103/PhysRevD.86.044009}{\emph{Phys. Rev. D}
  {\bfseries 86} (2012) 044009}
  [\href{https://arxiv.org/abs/1204.4593}{{\ttfamily 1204.4593}}].

\bibitem{Hohmann:2019nat}
M.~Hohmann, L.~J\"arv, M.~Kr\v{s}\v{s}\'ak and C.~Pfeifer, \emph{{Modified
  teleparallel theories of gravity in symmetric spacetimes}},
  \href{https://doi.org/10.1103/PhysRevD.100.084002}{\emph{Phys. Rev. D}
  {\bfseries 100} (2019) 084002}
  [\href{https://arxiv.org/abs/1901.05472}{{\ttfamily 1901.05472}}].

\bibitem{2013PASP..125..306F}
D.~{Foreman-Mackey}, D.W.~{Hogg}, D.~{Lang} and J.~{Goodman}, \emph{{emcee: The
  MCMC Hammer}}, \href{https://doi.org/10.1086/670067}{\emph{Publications of
  the Astronomical Society of the Pacific} {\bfseries 125} (2013) 306}
  [\href{https://arxiv.org/abs/1202.3665}{{\ttfamily 1202.3665}}].

\bibitem{2014RAA....14.1221Z}
C.~{Zhang}, H.~{Zhang}, S.~{Yuan}, S.~{Liu}, T.-J.~{Zhang} and Y.-C.~{Sun},
  \emph{{Four new observational H(z) data from luminous red galaxies in the
  Sloan Digital Sky Survey data release seven}},
  \href{https://doi.org/10.1088/1674-4527/14/10/002}{\emph{Research in
  Astronomy and Astrophysics} {\bfseries 14} (2014) 1221}
  [\href{https://arxiv.org/abs/1207.4541}{{\ttfamily 1207.4541}}].

\bibitem{Jimenez:2003iv}
R.~Jimenez, L.~Verde, T.~Treu and D.~Stern, \emph{{Constraints on the equation
  of state of dark energy and the Hubble constant from stellar ages and the
  CMB}}, \href{https://doi.org/10.1086/376595}{\emph{Astrophys. J.} {\bfseries
  593} (2003) 622} [\href{https://arxiv.org/abs/astro-ph/0302560}{{\ttfamily
  astro-ph/0302560}}].

\bibitem{Moresco:2016mzx}
M.~Moresco, L.~Pozzetti, A.~Cimatti, R.~Jimenez, C.~Maraston, L.~Verde et~al.,
  \emph{{A 6\% measurement of the Hubble parameter at $z\sim0.45$: direct
  evidence of the epoch of cosmic re-acceleration}},
  \href{https://doi.org/10.1088/1475-7516/2016/05/014}{\emph{JCAP} {\bfseries
  05} (2016) 014} [\href{https://arxiv.org/abs/1601.01701}{{\ttfamily
  1601.01701}}].

\bibitem{Simon:2004tf}
J.~Simon, L.~Verde and R.~Jimenez, \emph{{Constraints on the redshift
  dependence of the dark energy potential}},
  \href{https://doi.org/10.1103/PhysRevD.71.123001}{\emph{Phys. Rev. D}
  {\bfseries 71} (2005) 123001}
  [\href{https://arxiv.org/abs/astro-ph/0412269}{{\ttfamily
  astro-ph/0412269}}].

\bibitem{2012JCAP...08..006M}
M.~{Moresco}, A.~{Cimatti}, R.~{Jimenez}, L.~{Pozzetti}, G.~{Zamorani},
  M.~{Bolzonella} et~al., \emph{{Improved constraints on the expansion rate of
  the Universe up to z \raisebox{-0.5ex}\textasciitilde 1.1 from the
  spectroscopic evolution of cosmic chronometers}},
  \href{https://doi.org/10.1088/1475-7516/2012/08/006}{\emph{JCAP} {\bfseries
  2012} (2012) 006} [\href{https://arxiv.org/abs/1201.3609}{{\ttfamily
  1201.3609}}].

\bibitem{2010JCAP...02..008S}
D.~{Stern}, R.~{Jimenez}, L.~{Verde}, M.~{Kamionkowski} and S.A.~{Stanford},
  \emph{{Cosmic chronometers: constraining the equation of state of dark
  energy. I: H(z) measurements}},
  \href{https://doi.org/10.1088/1475-7516/2010/02/008}{\emph{JCAP} {\bfseries
  2010} (2010) 008} [\href{https://arxiv.org/abs/0907.3149}{{\ttfamily
  0907.3149}}].

\bibitem{Moresco:2015cya}
M.~Moresco, \emph{{Raising the bar: new constraints on the Hubble parameter
  with cosmic chronometers at z \ensuremath{\sim} 2}},
  \href{https://doi.org/10.1093/mnrasl/slv037}{\emph{Mon. Not. Roy. Astron.
  Soc.} {\bfseries 450} (2015) L16}
  [\href{https://arxiv.org/abs/1503.01116}{{\ttfamily 1503.01116}}].

\bibitem{Jimenez:2001gg}
R.~Jimenez and A.~Loeb, \emph{{Constraining cosmological parameters based on
  relative galaxy ages}},
  \href{https://doi.org/10.1086/340549}{\emph{Astrophys. J.} {\bfseries 573}
  (2002) 37} [\href{https://arxiv.org/abs/astro-ph/0106145}{{\ttfamily
  astro-ph/0106145}}].

\bibitem{SNLS:2011lii}
{\scshape SNLS} collaboration, \emph{{Supernova Constraints and Systematic
  Uncertainties from the First 3 Years of the Supernova Legacy Survey}},
  \href{https://doi.org/10.1088/0067-0049/192/1/1}{\emph{Astrophys. J. Suppl.}
  {\bfseries 192} (2011) 1} [\href{https://arxiv.org/abs/1104.1443}{{\ttfamily
  1104.1443}}].

\bibitem{Pan-STARRS1:2017jku}
{\scshape Pan-STARRS1} collaboration, \emph{{The Complete Light-curve Sample of
  Spectroscopically Confirmed SNe Ia from Pan-STARRS1 and Cosmological
  Constraints from the Combined Pantheon Sample}},
  \href{https://doi.org/10.3847/1538-4357/aab9bb}{\emph{Astrophys. J.}
  {\bfseries 859} (2018) 101}
  [\href{https://arxiv.org/abs/1710.00845}{{\ttfamily 1710.00845}}].

\bibitem{Brout:2022vxf}
D.~Brout et~al., \emph{{The Pantheon+ Analysis: Cosmological Constraints}},
  \href{https://doi.org/10.3847/1538-4357/ac8e04}{\emph{Astrophys. J.}
  {\bfseries 938} (2022) 110}
  [\href{https://arxiv.org/abs/2202.04077}{{\ttfamily 2202.04077}}].

\bibitem{Ross:2014qpa}
A.J.~Ross, L.~Samushia, C.~Howlett, W.J.~Percival, A.~Burden and M.~Manera,
  \emph{{The clustering of the SDSS DR7 main Galaxy sample \textendash{} I. A 4
  per cent distance measure at $z = 0.15$}},
  \href{https://doi.org/10.1093/mnras/stv154}{\emph{Mon. Not. Roy. Astron.
  Soc.} {\bfseries 449} (2015) 835}
  [\href{https://arxiv.org/abs/1409.3242}{{\ttfamily 1409.3242}}].

\bibitem{2011MNRAS.416.3017B}
F.~{Beutler}, C.~{Blake}, M.~{Colless}, D.H.~{Jones}, L.~{Staveley-Smith},
  L.~{Campbell} et~al., \emph{{The 6dF Galaxy Survey: baryon acoustic
  oscillations and the local Hubble constant}},
  \href{https://doi.org/10.1111/j.1365-2966.2011.19250.x}{\emph{Monthly Notices
  of the Royal Astronomical Society} {\bfseries 416} (2011) 3017}
  [\href{https://arxiv.org/abs/1106.3366}{{\ttfamily 1106.3366}}].

\bibitem{Bourboux:2017cbm}
H.~du~Mas~des Bourboux et~al., \emph{{Baryon acoustic oscillations from the
  complete SDSS-III Ly$\alpha$-quasar cross-correlation function at $z=2.4$}},
  \href{https://doi.org/10.1051/0004-6361/201731731}{\emph{Astron. Astrophys.}
  {\bfseries 608} (2017) A130}
  [\href{https://arxiv.org/abs/1708.02225}{{\ttfamily 1708.02225}}].

\bibitem{Zhao:2018gvb}
G.-B.~Zhao et~al., \emph{{The clustering of the SDSS-IV extended Baryon
  Oscillation Spectroscopic Survey DR14 quasar sample: a tomographic
  measurement of cosmic structure growth and expansion rate based on optimal
  redshift weights}}, \href{https://doi.org/10.1093/mnras/sty2845}{\emph{Mon.
  Not. Roy. Astron. Soc.} {\bfseries 482} (2019) 3497}
  [\href{https://arxiv.org/abs/1801.03043}{{\ttfamily 1801.03043}}].

\bibitem{Alam:2016hwk}
{\scshape BOSS} collaboration, \emph{{The clustering of galaxies in the
  completed SDSS-III Baryon Oscillation Spectroscopic Survey: cosmological
  analysis of the DR12 galaxy sample}},
  \href{https://doi.org/10.1093/mnras/stx721}{\emph{Mon. Not. Roy. Astron.
  Soc.} {\bfseries 470} (2017) 2617}
  [\href{https://arxiv.org/abs/1607.03155}{{\ttfamily 1607.03155}}].

\bibitem{Planck:2018vyg}
{\scshape Planck} collaboration, \emph{{Planck 2018 results. VI. Cosmological
  parameters}},
  \href{https://doi.org/10.1051/0004-6361/201833910}{\emph{Astron. Astrophys.}
  {\bfseries 641} (2020) A6}
  [\href{https://arxiv.org/abs/1807.06209}{{\ttfamily 1807.06209}}].

\bibitem{2009ApJ...707..916F}
D.J.~{Fixsen}, \emph{{The Temperature of the Cosmic Microwave Background}},
  \href{https://doi.org/10.1088/0004-637X/707/2/916}{\emph{The Astrophysical
  Journal} {\bfseries 707} (2009) 916}
  [\href{https://arxiv.org/abs/0911.1955}{{\ttfamily 0911.1955}}].

\bibitem{Linder:2009jz}
E.V.~Linder, \emph{{Exponential Gravity}},
  \href{https://doi.org/10.1103/PhysRevD.80.123528}{\emph{Phys. Rev. D}
  {\bfseries 80} (2009) 123528}
  [\href{https://arxiv.org/abs/0905.2962}{{\ttfamily 0905.2962}}].

\end{thebibliography}\endgroup

\end{document}